\documentclass[english,dvipsnames]{article}
\usepackage{lmodern}

\usepackage[T1]{fontenc}
\usepackage[latin9]{inputenc}
\usepackage{geometry}
\usepackage{color}
\usepackage{babel}
\usepackage{float}
\usepackage{url}
\usepackage{dsfont}
\usepackage{amsmath}
\usepackage{amsthm}
\usepackage{amssymb}
\usepackage{stmaryrd}
\usepackage{graphicx}
\usepackage{setspace}
\usepackage[authoryear]{natbib}
\usepackage[unicode=true,bookmarks=true,bookmarksnumbered=false,bookmarksopen=false,
 breaklinks=false,pdfborder={0 0 1},backref=page,colorlinks=true]
 {hyperref}
\hypersetup{pdftitle={Unbiased MCMC},
 linkcolor=blue,urlcolor=blue,citecolor=blue}

\numberwithin{equation}{section}
\theoremstyle{plain}
\newtheorem{theorem}{Theorem}[section]
\theoremstyle{plain}
\newtheorem{assumption}{Assumption}
\usepackage{listings}
\usepackage{color}
\definecolor{dkgreen}{rgb}{0,0.6,0}
\definecolor{gray}{rgb}{0.5,0.5,0.5}
\definecolor{mauve}{rgb}{0.58,0,0.82}

\usepackage{algorithm}

\begin{document}

\title{Unbiased Markov Chain Monte Carlo: what, why, and how\footnote{A version of this document will appear
as a chapter in the 2nd edition of the Handbook of MCMC.}}
\author{Yves F. Atchad\'e (Boston University),
Pierre E. Jacob (ESSEC Business
School)}

\maketitle

\begin{abstract}
  This document presents methods to remove the \emph{initialization} or
\emph{burn-in} bias from Markov chain Monte Carlo (MCMC) estimates, with consequences on parallel computing,
convergence diagnostics and performance assessment. The document is written as an introduction to these methods for MCMC users. Some theoretical results are mentioned, but the focus is on the methodology.
\end{abstract}

\section{Introduction\label{sec:introduction}}

\subsection{Initialization bias in MCMC\label{sec:introduction:1}}

The object of interest is a probability measure $\pi$ on a space
$(\mathbb{X},\mathcal{X})$. An MCMC algorithm generates a chain
$(X_{t})_{t\geq0}$ via a $\pi$-invariant Markov transition kernel $P$, starting
from an initial distribution $\pi_{0}$ which is not equal to $\pi$.  The
marginal distribution of $X_t$ at time $t$ is denoted by $\pi_t$. Chains generated by MCMC
algorithms are often provably ergodic, i.e. 
\begin{equation} 
  |\pi_t - \pi|_{\text{TV}} \leq b(t) \to 0 \text{ as } t\to\infty,\label{eq:ergodicity}
\end{equation} 
for some decreasing function $b$, where $|\cdot|_{\text{TV}}$ denotes the total variation (TV). Recall that for two 
probability measures $\mu$ and $\nu$ on $(\mathbb{X},\mathcal{X})$, $|\mu - \nu|_{\text{TV}} = \sup_{A\in\mathcal{X}}|\mu(A) - \nu(A)|$.
Results of the
form \eqref{eq:ergodicity} abound in the literature \citep[see e.g. bibliographical notes in Chapter 15 in][]{douc2018MarkovChains}, but the function $b(t)$
typically features unspecified quantities, and thus cannot actually be
evaluated for a given iteration $t$. There are exceptions, e.g. Theorem 11 in
\citet{rosenthal1995minorization},
or the references in Section 3.5 in \citet{roberts2004general}, where bounds are made fully explicit for non-trivial MCMC algorithms,
using both analytical and numerical computation. 
Efforts have been long made to design numerical recipes that would provide explicit upper bounds
as automatically as possible \citep[e.g.][]{johnson1996studying,cowles1998simulation,johnson1998coupling} and 
unbiased MCMC methods contribute to that effort (more on this in Section \ref{sec:byproducts:ubounds}).

The initialization bias comes from the marginal distribution $\pi_t$, at any
time $t$, being different from $\pi$.  There may be other sources of bias in MCMC, such as the use of pseudo-random rather than random numbers,
limited floating point precision, as well as various deliberate approximations that can accelerate computation.
This document focuses on initialization bias.
We introduce a test function $h$ in
$L^{p}(\pi)=\{f:\pi(\left|f\right|^{p})<\infty\}$ where $\pi(f):=\int
f(x)\pi({\rm d}x)$ and $p\geq 1$. The MCMC estimator of $\pi(h)$ is the ergodic average
$t^{-1}\sum_{s=0}^{t-1}h(X_{s})$, possibly after discarding an initial portion
of the trajectory.  The initialization bias is defined as
$\mathbb{E}[t^{-1}\sum_{s=0}^{t-1}h(X_{s})] -  \pi(h)$. It is only zero if
$\pi_0$ is precisely $\pi$; most often it is considered unknown.  The bias
vanishes as $t\to\infty$ and becomes a negligible part of the mean squared
error, which is dominated by the variance $v(P,h)$ in
the Central Limit Theorem (CLT): \begin{equation}
\sqrt{t}\left(\frac{1}{t}\sum_{s=0}^{t-1}h(X_{s})-\pi(h)\right)\overset{d}{\to}\text{Normal}(0,v(P,h)),\qquad\text{as
}t\to\infty.\label{eq:clt} \end{equation}
We will recall below some conditions under which the CLT holds (see Assumption \ref{asmpt:meetingmoments}).

\begin{figure}
  \begin{center}
    {\includegraphics[width=\textwidth]{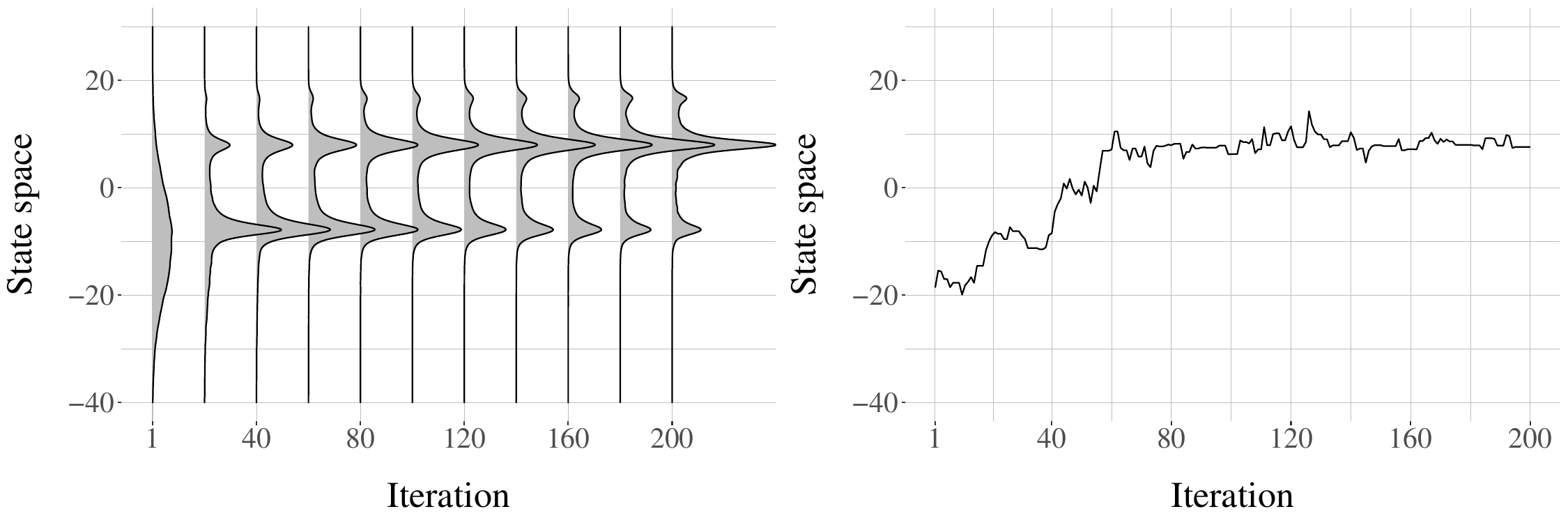}} 
    \caption{Convergence of the marginal distribution $\pi_t$ to $\pi$ (left) and a realized trajectory (right), here generated
	    by a Metropolis--Rosenbluth--Teller--Hastings (MRTH) algorithm with Normal random walk proposals. The target distribution 
	    is described in Example 3.1 of \citet{robert1995convergence}. This setting is used for all 
	    illustrations of this document.\label{fig:marginalconvergence}
    }
  \end{center}
\end{figure}  

Despite its asymptotic disappearance, the initialization bias poses practical
issues. The bias is an obstacle to the parallelization of MCMC
computation \citep{rosenthal2000parallel}.  Indeed, 
users can generate short MCMC runs independently in parallel, but the bias 
prevents the consistent estimation of $\pi(h)$
by averages over the independent runs. The bias can be
reduced by discarding a larger initial portion of each parallel run, but
the choice of the length to discard is both difficult and critical. 
This is in part because a sufficient length for the bias to be small 
is vastly different from one application to the next; 
numbers of iterations reported in the literature span many orders of magnitude,
e.g. \citet{metropolis1953equation} perform a few dozen sweeps whereas \citet{mccartan2020sequential}
mention a run of a trillion ($10^{12}$) MCMC iterations for the task of sampling redistricting plans.

\subsection{The promise of unbiased MCMC\label{sec:introduction:2}}

\emph{Unbiased MCMC} removes the
initialization bias.  The key requirement of these methods is that the user
can generate certain couplings of Markov chains (see 
Section \ref{sec:introduction:3}).  This requirement is weaker than that of Coupling
From The Past \citep{propp1996exact}, and thus unbiased MCMC is more widely
applicable; but it does not provide perfect samples from $\pi$.  Instead, these
methods pioneered by \citet{glynn2014exact} generate unbiased approximations of $\pi$ in the form of signed
empirical measures:
\begin{equation}
  \hat{\pi} = \sum_{n=1}^N \omega_n \delta_{Z_n},
  \label{eq:signedmeas}
\end{equation}
where $N$ is a random integer, $(Z_n)_{n=1}^N$ are atoms on the state space $\mathbb{X}$,
and $(\omega_n)_{n=1}^N$ are real-valued weights (see Section \ref{sec:unbiasedmcmc:signedmeasure} for the precise
construction). The lack-of-bias
property of $\hat{\pi}$ means that, for a class of functions $h$, $\hat{\pi}(h) := \sum_{n=1}^N \omega_n h(Z_n)$
has expectation exactly equal to $\pi(h)$.
 The significance of the lack of bias
is that users can generate independent copies of $\hat{\pi}(h)$ 
in parallel and average over the copies to obtain consistent estimators
of $\pi(h)$, converging at the standard Monte Carlo rate if the estimators have a finite variance.
Theorem \ref{thm:finitepmoments} below provides conditions under which this holds.
The lack of bias is thus clearly appealing as a means 
toward a parallel-friendly consistent Monte Carlo scheme; 
there are other appeals discussed in Section \ref{sec:discussion}.

With unbiased signed measures, the question of initialization bias seems to be
resolved. However, both the computing time and
the variance can be prohibitively high. To quantify
the price of removing the bias, for any function $h$, we can compute the \emph{inefficiency},
a key descriptor
of asymptotic performance for unbiased estimators \citep{glynn1992asymptotic},
defined as
the expected computing cost multiplied by the variance of $\hat{\pi}(h)$.
Both the expected cost and the variance can be estimated from independent runs.
Meanwhile, the standard MCMC estimator has a cost proportional to the number of iterations $t$,
and a variance of order $v(P,h)/t$ as $t\to\infty$,
with $v(P,h)$ the asymptotic variance in the CLT \eqref{eq:clt}. Thus, the asymptotic inefficiency
of MCMC is measured by $v(P,h)$.
The first unbiased estimators constructed from coupled chains in
\citet{glynn2014exact}, with MCMC applications presented in
\citet{agapiou2018unbiased}, were not always competitive with MCMC
in terms of asymptotic inefficiency. The simple enhancements proposed in
\citet{jacob2018smoothing,joa2020,VanettiDoucet2020} (see Section
\ref{sec:unbiasedmcmc:unbiasedmcmc}) led to unbiased MCMC estimators that are
nearly as efficient as standard MCMC estimators in a wide range of settings.

\subsection{Successful couplings of Markov chains\label{sec:introduction:3}}

Unbiased MCMC belongs to a family of algorithms that require
couplings of Markov chains, as in Coupling From The Past
(CFTP, see \citet{propp1996exact}), circularly-coupled MCMC
\citep{neal1999circularly}, and Johnson's convergence diagnostics
\citep{johnson1996studying,johnson1998coupling}.  A \emph{coupling} of two
distributions $p$ and $q$ on $\mathbb{X}$ refers to a joint distribution on
$\mathbb{X}\times \mathbb{X}$, with prescribed marginals $p$ and $q$. For
Markov chains, a coupling refers to a joint process $(X_t,Y_t)$ such that
$(X_t)$ and $(Y_t)$ are individually identical to prescribed Markov chains.
For simplicity, we focus on Markovian couplings, where $(X_t,Y_t)$ itself forms a
Markov chain, with initial distribution $\bar{\pi}_0$ on $\mathbb{X}\times\mathbb{X}$ and Markov transition
$\bar{P}$.

\begin{algorithm}
  \begin{enumerate}
    \item Sample $(X_0,Y_0)$ from $\bar{\pi}_0$, .
    \item If $L\geq1$, for $t=1,\ldots,L$, sample $X_{t}$ from $P(X_{t-1},\cdot)$.
    \item For $t\geq L$, sample $(X_{t+1},Y_{t-L+1})$ from $\bar{P}((X_{t},Y_{t-L}),\cdot)$
      until $X_{t+1}=Y_{t-L+1}$ and $t+1\geq \ell$.
  \end{enumerate}
  \caption{Successful coupling of chains with lag $L$ and length $\ell$. Coupled initial distribution: $\bar{\pi}_0$, 
    transition: $P$, coupled transition: $\bar{P}$,
  meeting time: $\tau = \inf\{t\geq L: X_{t}=Y_{t-L}\}$. \label{alg:laggedchains}}
  \end{algorithm}

\begin{figure}
\begin{center}
  {\includegraphics[width=\textwidth]{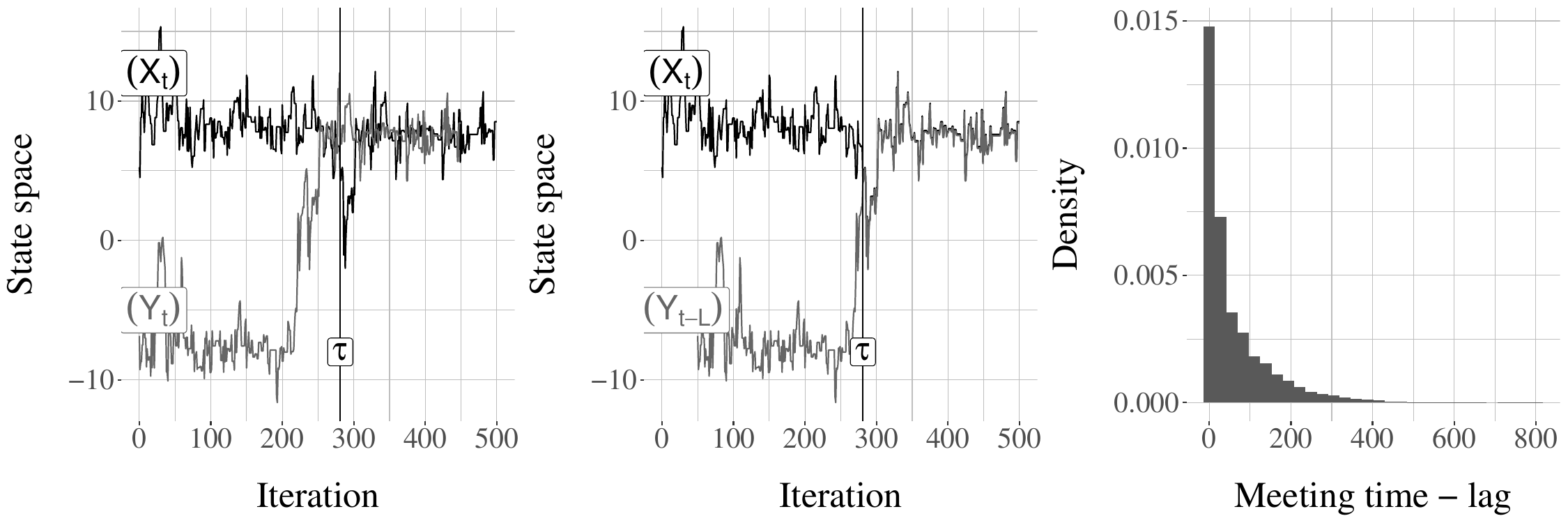}} 
  \caption{Successful coupling of Markov chains, meeting at time $\tau$ with a lag $L$:
    $X_t = Y_{t-L}$ for all $t\geq \tau$. Left: trajectory of $(X_t,Y_t)$.
    Middle: trajectory of $(X_t,Y_{t-L})$. Right: histogram of $10^4$ independent copies of $\tau-L$.
\label{fig:laggedchains}}
\end{center}
\end{figure}

Unbiased MCMC requires draws of $(X_t,Y_t)$ such that both chains $(X_t)$ and
$(Y_t)$ are copies of the same Markov process that has initial distribution $\pi_0$, and transition $P$. Thus, $\bar{\pi}_0$ should be a coupling of $\pi_0$ with itself: $\bar{\pi}_0(A\times \mathbb{X})$ and $\bar{\pi}_0(\mathbb{X}\times A)$ should equal $\pi_0(A)$ for any measurable set $A$. 
The transition $\bar{P}$ should be a coupling of $P$ with itself: $\bar{P}((x,y), A\times \mathbb{X})$ and $\bar{P}((x,y), \mathbb{X}\times A)$ should equal $P(x,A)$ and $P(y,A)$ respectively for all $A$. Furthermore, for each
trajectory of $(X_t,Y_t)$ we require the existence of a finite
\emph{meeting time} $\tau$ such that $X_t = Y_{t-L}$ for all $t\geq \tau$,
where $L\geq 1$ is a user-chosen \emph{lag} parameter.  Coupling resulting in
finite meeting times are called \emph{successful} in this document, following
\citet{pitman1976coupling}. The construction is described in Algorithm \ref{alg:laggedchains}.
A realization of a successful coupling is shown in
Figure \ref{fig:laggedchains}, where $L=50$ and $\ell = 500$.

The key assumption in this document is about the tails of the meeting times associated
with coupled chains started from an independent coupling of $\pi$ and $\pi$, and propagated
according to $\bar{P}$. Thus, the assumption does not involve $\pi_0$. 
The assumption is taken from \citet{douc2022solving} and is equivalent to $\mathbb{P}(\tau > t)$ decaying at a polynomial rate as $t\to\infty$.

\begin{assumption}
  There exists $\kappa\geq 1$ such that, if two chains $(X_t)$ and $(Y_t)$ start from $\pi$ independently,
  and evolve according to $\bar{P}$, the meeting time $\tau=\inf\{t\geq 1: X_t = Y_t\}$
  satisfies $\mathbb{E}[\tau^\kappa]<\infty$. \label{asmpt:meetingmoments}
\end{assumption}

The assumption has implications on the marginal chain. For example, it implies that the CLT in \eqref{eq:clt} holds for any $h\in L^p(\pi)$ with $p>2\kappa/(\kappa-1)$
\citep{douc2022solving}. The assumption is useful to study unbiased MCMC estimators
as illustrated in Theorems \ref{thm:finitepmoments} and \ref{thm:cltunbiasedMCMC} below.

Assumption \ref{asmpt:meetingmoments} can be verified in different ways, on a case-by-case basis.
If the
transition $P$ satisfies a Lyapunov drift condition with function $V$, 
and if $\bar{P}$ results in a non-zero probability of meeting over one step when both chains are
simultaneously in a level set of $V$, then Section 3.2 in \citet{joa2020} provides
a result on the tails of $\tau$.
That result applies in the numerous cases where explicit Lyapunov functions have already been elicited: 
random walk Metropolis--Rosenbluth--Teller--Hastings,
abbreviated MRTH \citep[see][]{roberts1996geometric}, Langevin Monte Carlo
\citep{durmus2022geometric}, Hamiltonian Monte Carlo
\citep{durmus2017convergence}, and many examples of Gibbs samplers.  A
similar result applies under polynomial drift conditions \citep[Section 1.4
in][]{middleton2020unbiased}.

As a concrete example, a simple random walk MRTH algorithm
is implemented in \texttt{R} in Figure \ref{fig:Rcode}, where the function \texttt{U} represents $x\mapsto - \log \pi(x)$ up to an additive constant. It employs a coupling
of Normal proposal distributions presented in Section 2.3 of \citet{bou2020coupling},
and in Section \ref{sec:designcoupling:ingredients} below.
This coupling was employed to generate all figures in this document.
Assumption \ref{asmpt:meetingmoments}
can then be verified for all $\kappa>1$ via Proposition 4 in \citet{joa2020}
using the geometric drift function $V(x) = \pi(x)^{-1/2}$ under the conditions of Theorem 3.2 in \citet{roberts1996geometric}.
Section \ref{sec:designcoupling} provides more discussion on the design of
successful couplings of MCMC algorithms.

\begin{figure}[ht!]
  \centering
\begin{lstlisting}
# Metropolis-Rosenbluth-Teller-Hastings transition with Normal proposals
mrth = function(x, U, sigma)
{
  # proposal = current location + Normal(0,sigma^2)
  xprop = x + sigma * rnorm(length(x))
  # log Uniform to accept/reject proposals
  logu = log(runif(1))
  # return state according to decision to accept or not
  return(if (logu < (U(x) - U(xprop))) xprop else x)
}
# coupling of MRTH transition with Normal proposals
coupledmrth = function(x, y, U, sigma)
{
  # draw proposals using maximal coupling of Bou-Rabee, Eberle and Zimmer (AAP 2020)
  xstd = rnorm(length(x)) # standard Normal variables
  z = (x - y) / sigma # length(sigma) could be 1 or length(x)
  e = z / sqrt(sum(z^2)) # normalise
  logu = log(runif(1)) 
  sameprop = (logu < sum(dnorm(xstd + z, log = TRUE) - dnorm(xstd, log = TRUE)))
  ystd = if (sameprop) xstd + z else xstd - 2 * sum(e * xstd) * e
  # xprop is marginally Normal(x,sigma^2) 
  xprop = x + sigma * xstd
  # yprop is marginally Normal(y,sigma^2)
  yprop = y + sigma * ystd
  # log Uniform to accept/reject proposals
  logu = log(runif(1))
  # decision to accept or not
  xaccept = (logu < (U(x) - U(xprop)))
  yaccept = (logu < (U(y) - U(yprop)))
  # return state according to decision
  return(list(nextx = if (xaccept) xprop else x,
              nexty = if (yaccept) yprop else y,
              nextxequalsnexty = sameprop && xaccept && yaccept))
}
\end{lstlisting}
\caption{\texttt{R} code for MRTH algorithm with Normal random walk proposals,
  and a coupling of it. This defines the transition $P$ and coupled transition $\bar{P}$ required by Algorithm \ref{alg:laggedchains}.
  Inputs: current states \texttt{x} and \texttt{y}, a potential function \texttt{U} corresponding to $x\mapsto -\log\pi(x)$, proposal standard deviation \texttt{sigma} (a scalar or a vector of the same length as \texttt{x} and \texttt{y}).}
  \label{fig:Rcode}
\end{figure}

\section{Unbiased MCMC \label{sec:unbiasedmcmc}}

This section presents unbiased MCMC estimators, 
assuming that successful couplings can be implemented and deferring to Section \ref{sec:designcoupling} for more on the design of such couplings.
We start with 
bias removal techniques in Section \ref{sec:unbiasedmcmc:telescop},
re-derive unbiased MCMC via the Poisson equation in Section \ref{sec:unbiasedmcmc:poisson},
and present more efficient versions in Sections \ref{sec:unbiasedmcmc:unbiasedmcmc}-\ref{sec:unbiasedmcmc:signedmeasure}.
In Section \ref{sec:unbiasedmcmc:efficiency} we comment on performance, cost, parallel computing and tuning.

\subsection{Bias removal with a telescope\label{sec:unbiasedmcmc:telescop}}

\textbf{Randomized telescoping sums.} Consider a quantity of interest expressed
as the limit $b_\infty$ of a deterministic sequence $(b_k)_{k\geq 0}$
as $k\to \infty$. Defining $b_{-1} = 0$
and $a_k = b_{k}-b_{k-1}$ for $k\geq 0$,
we can write $b_\infty$ as
the series $\sum_{k=0}^\infty a_k$. 
Assume that the time to compute the $k$-th term in the sequence
$(b_k)_{k\geq 0}$ is equal to $k$.  Can we
estimate a series $\sum_{k=0}^\infty a_k$ without bias in finite time?  The following
reasoning appears in \citet{glynn1983,rychlik1990unbiased}.  Let
$\xi$ be a random variable on $\{0,1,2\ldots\}$ with $p_k:= \mathbb{P}(\xi =
k)>0$ for all $k\geq 0$, called the \emph{truncation variable}.  Then
sample $\xi$ and compute: $G = {a_\xi}/{p_\xi}$.  If the expectation of $|G|$ is
finite then $\mathbb{E}[G] = b_\infty$, and its expected cost is
$\mathbb{E}[\xi] = \sum_{k\geq 0} k\, p_k$. The cost is smaller if $(p_k)$ decay
faster.  However, the variance of $G$ involves $\mathbb{E}[G^2] = \sum_{k\geq 0}
a_k^2/p_k$ which is smaller if $(p_k)$ decay slower. 
Thus, the estimator $G$ can only have finite expected cost and finite variance if $(b_k)$ converges fast enough
and if $(p_k)$ is chosen adequately.
An alternative is to
sample $\xi$ and then compute $H =  a_0 + \sum_{k=1}^\xi a_k/\mathbb{P}(\xi\geq
k)$.  The estimator $H$ also has expectation $b_\infty$, its cost
is similar to that of $G$, but its variance is finite under weaker conditions
than that of $G$. The estimators $G$ and $H$ are termed \emph{single term} and
\emph{coupled sum}, and are discussed in detail in \citet{rhee2015unbiased,vihola2018unbiased}.

\textbf{Bias removal in MCMC.} Direct use of the above strategy to remove the
bias of MCMC averages, where $b_k = k^{-1}\sum_{s=0}^{k-1} h(X_s)$, is considered
in \citet{mcleish2011general}. An immediate difficulty is that ergodic averages
converge at the (slow) Monte Carlo rate, resulting in unbiased estimators that
tend to have either a large cost or a large variance. The convergence of marginal
distributions $\pi_t \to \pi$ e.g. in total variation is comparably faster.
Using contractive couplings of
Markov chains and exploiting the (fast) convergence of marginal distributions,
\citet{glynn2014exact} propose a debiasing strategy with a truncation variable that determines the length of the coupled chains. \citet{glynn2014exact} also consider
the case where one could sample from a measure that minorizes the Markov transition. In that case their coupling
results in pairs of chains that meet exactly, thus providing a natural stopping criterion for the coupled chains, and removing the need for truncation variables. 
\citet{agapiou2018unbiased}
employ contractive couplings of certain MCMC algorithms to obtain unbiased estimators and describe the practical choices associated
with the specification of the truncation variable. \citet{jacob2018smoothing} find that the conditional
particle filter, which is an MCMC algorithm for continuous state space models
\citep{andrieu2010particle}, could be coupled such that a pair of chains would
meet, often quickly, without the need for a truncation variable.  \citet{joa2020} find that many MCMC
algorithms can be coupled in that way, i.e. successfully, using  maximal couplings as in \citet{johnson1998coupling}. 
In this document we will focus on unbiased estimators obtained from successful couplings, where chains meet exactly, and truncation variables are not required.

\textbf{First unbiased MCMC estimator.}
The idea of \citet{glynn2014exact} in the context of successful couplings introduced in Section 
\ref{sec:introduction:3} goes as follows.
Write $\pi(h)$ as a telescopic sum, for all $k\geq 0$, for any choice of lag $L$,
\begin{equation}
	\pi(h) =   \lim_{t\to\infty } \pi_t(h) = \pi_{k}(h) + \sum_{j=1}^\infty \pi_{k+jL}(h) - \pi_{k+(j-1)L}(h).
	\label{eq:telescopictrick}
\end{equation}
For all $t\geq 0$, $X_{t}$ and $Y_t$ have the same distribution $\pi_t$, 
thus $\pi_{k+jL}(h) = \mathbb{E}[h(X_{k+jL})]$ and $\pi_{k+(j-1)L}(h)=\mathbb{E}[h(Y_{k+(j-1)L})]$.
A swap of expectation and limit suggests that 
\begin{equation}
	H_{k}:=h(X_{k}) + \sum_{j=1}^\infty (h(X_{k+jL}) - h(Y_{k+(j-1)L})),
	\label{eq:HkL}
\end{equation}
is an unbiased estimator of $\pi(h)$. 
For instance, if $|h|\leq 1$
then $\sum_{j=1}^\infty |h(X_{k+jL}) - h(Y_{k+(j-1)L})| \leq 2\max(0,(\tau - k)/L)$.
Therefore, if Assumption \ref{asmpt:meetingmoments} holds with $\kappa=1$, 
then by Fubini's 
theorem $H_k$ indeed satisfies $\mathbb{E}[H_k] = \pi(h)$. Higher moments of $H_k$ can be controlled as in Theorem \ref{thm:finitepmoments} below.
The infinite sum in \eqref{eq:HkL} can be computed in finite time since 
the differences $h(X_{k+jL}) - h(Y_{k+(j-1)L})$ are equal to zero for all $j$ such that $k+jL \geq \tau$.

\subsection{Alternative construction via the Poisson equation\label{sec:unbiasedmcmc:poisson}}

\textbf{Poisson equation and bias.} \citet{douc2022solving} provide an alternative derivation of $H_k$
in \eqref{eq:HkL}
via the Poisson equation.
Write $Pf(x) = \int P(x,dx')f(x')$ for a function $f:\mathbb{X}\to\mathbb{R}$.
  A function $g$ in $L^{1}(\pi)$ is a solution of the
  Poisson equation associated with $h$ and $P$ if 
  \begin{equation} \label{eq:poisson}
	  g(x)-Pg(x)=h(x)-\pi(h)  \quad \forall x\in\mathbb{X}.
  \end{equation}
For example, the function
\begin{equation}\label{eq:starfish}
    g_\star: x\mapsto \sum_{t=0}^{\infty} P^{t}\left\{h-\pi(h)\right\}(x),
  \end{equation}
  may be a solution to \eqref{eq:poisson} if it is well-defined; see Chapter 21 of \citet{douc2018MarkovChains}.
  As noted in \citet{douc2022solving}, if $h\in L^p(\pi)$ and if $\kappa$ in Assumption \ref{asmpt:meetingmoments} is such that
  $p>2\kappa/(\kappa-1)$, then $g_\star$ is indeed in $L^1(\pi)$. In that case, all solutions to \eqref{eq:poisson} are equal to $g_\star$ up to an additive constant.
  It is known that \eqref{eq:starfish} is related to MCMC bias.
  Indeed, consider the ergodic average $t^{-1}\sum_{s=0}^{t-1}h(X_{s})$ when the chain starts from a fixed $x_0\in\mathbb{X}$.
  The bias is $\mathbb{E}_{x_{0}}[t^{-1}\sum_{s=0}^{t-1}h(X_{s})]-\pi(h)$, and
  if we multiply by $t$ and consider the limit $t\to\infty$,
\begin{equation}
\lim_{t\to\infty}t\times \{ \mathbb{E}_{x_{0}}[t^{-1}\sum_{s=0}^{t-1}h(X_{s})]-\pi(h)\} =
\lim_{t\to\infty} \sum_{s=0}^{t-1}\mathbb{E}_{x_{0}}[h(X_{s})-\pi(h)] = g_\star(x_{0}),
  \label{eq:fishyasymptbias}
\end{equation}
with $g_\star$ as in \eqref{eq:starfish}, and $\mathbb{E}_{x_0}$ referring to expectations with respect to the chain started from $X_0 = x_0$ \citep{kontoyiannis2009notes}.

\textbf{Estimation of $g$ and unbiased MCMC.}
Consider the function
$x\mapsto g(x,y):= g_\star(x)-g_\star(y)$, equal to $g_\star$ up to the constant
$g_\star(y)$, for any fixed $y\in\mathbb{X}$, and thus solution of the Poisson equation under the aforementioned conditions.  This solution can be
written 
\begin{equation}
g(x,y)=\sum_{t\geq 0}\{P^t h(x)-P^t h(y)\}.
\label{eq:gxy}\end{equation}
Consider
chains coupled successfully with no lag ($L=0$), started from $X_0 = x$ and $Y_0 = y$.
Then $h(X_t)$ and $h(Y_t)$ have expectation equal to $P^th(x)$ and $P^th(y)$
for all $t\geq 0$, but for $t$ larger than $\inf\{t\geq 1: X_t = Y_t\}$ we have
$h(X_t)-h(Y_t)=0$.  This suggests the following unbiased estimator of
$g(x,y)$:
\begin{equation}
G(x,y):=\sum_{t=0}^{\tau-1}\{h(X_{t})-h(Y_{t})\},
  \label{eq:Gxy}
\end{equation}
where here $\tau=\inf\{t\geq 1: X_t = Y_t\}$.  With the
ability to estimate solutions of the Poisson equation we might envision the
estimation of $\pi(h)$ via the re-arranged equation: $\pi(h) = h(x) + Pg(x) -
g(x)$.  Setting $x\in\mathbb{X}$ arbitrarily, sample $X_1 \sim P(x,\cdot)$
(performing one step of MCMC), and sample $G(X_1,x)$ (running two chains,
initialized at $X_1$ and $x$, until they meet).  Then $\mathbb{E}_x[G(X_1,x)] =
Pg_\star(x)-g_\star(x)$, therefore $h(x) + G(X_1,x)$ is an unbiased
estimator of $\pi(h)$.  It is in fact exactly $H_{k}$ in
\eqref{eq:HkL} with $k=0$, $L=1$ and $\pi_0 = \delta_x$.

\subsection{Unbiased MCMC estimators\label{sec:unbiasedmcmc:unbiasedmcmc}}

\textbf{Improved efficiency by averaging.} Simple modifications of \eqref{eq:HkL}
can go a long way to improve its efficiency,
as noted in \citet{jacob2018smoothing,joa2020}.
Consider a run of Algorithm \ref{alg:laggedchains} with lag $L\geq 1$ \citep{VanettiDoucet2020} 
and length $\ell\geq 0$, from which we can construct unbiased
estimators $H_{k},\ldots,H_{\ell}$ as in \eqref{eq:HkL}
for a range of integers $k,\ldots,\ell$ where $0\leq k\leq \ell$. Since these estimators $(H_{t})_{t=k}^{\ell}$
are unbiased, their average is unbiased as well. We define
\begin{equation}
  H_{k:\ell}=\frac{1}{\ell-k+1}\sum_{t=k}^{\ell} H_t.\label{eq:H_kellL_basic}
\end{equation}
After some algebraic manipulations,
that estimator reads
\begin{equation}
  H_{k:\ell}=\underbrace{\frac{1}{\ell-k+1}\sum_{t=k}^{\ell}h(X_{t})}_{\text{MCMC}}
+\underbrace{\sum_{t=k+L}^{\tau-1} v_t(k,\ell,L)
\left\{ h(X_{t})-h(Y_{t-L})\right\}}_{\text{bias cancellation}},\label{eq:H_kellL}
\end{equation}
with weights $ v_t(k,\ell,L)$ defined precisely below. In \eqref{eq:H_kellL} the sum $\sum_{t=k+L}^{\tau - 1}$ is zero if $k+L\geq\tau$.
The first term on the right-hand side is the regular MCMC ergodic average,
computed from the trajectory $(X_k,\ldots,X_\ell)$.
The second term performs bias cancellation from weighted differences between the chains. 
The weight $v_t(k,\ell,L)$ is defined
as the number of appearances of the difference $h(X_t)-h(Y_{t-L})$
in the bias cancellation terms of $H_{k},\ldots,H_{\ell}$,
divided by $\ell-k+1$. Note that, for two positive integers $a\leq b$, the number of multiples of $L$ within $\{a,\ldots,b\}$
equals $\lfloor b/L \rfloor - \lceil a/L \rceil + 1$.
Using this and routine calculations, the weight can be written as 
\begin{equation}
v_t(k,\ell,L) = \frac{\lfloor(t-k) / L\rfloor - \lceil \max(L, t-\ell)/L\rceil + 1}{\ell-k+1}.
  \label{eq:vtweight}
\end{equation}
Both the number of terms in the bias cancellation and their weights can 
be reduced by increasing the tuning parameters $k,\ell,L$. This impacts the cost of obtaining $H_{k:\ell}$:
indeed, if we count the cost of sampling from the MCMC transition $P$ as one unit, and
the cost of sampling from the coupled transition $\bar{P}$ as one unit if the
chains have already met and two units if they have not, then 
\begin{equation} \label{eq:costHkell}
  \text{cost}(H_{k:\ell}) =  L +2(\tau-L) + \max(0,\ell-\tau).
\end{equation}
Section \ref{sec:unbiasedmcmc:efficiency} proposes guidance on tuning
unbiased MCMC, and Figure \ref{fig:ctimeinef} provides an illustration of the effect of tuning. 

\textbf{Variance reduction.} The estimator in \eqref{eq:H_kellL}
is presented with a generic lag $L$ following the observation of \citet{VanettiDoucet2020}
that increasing $L$ can lead to significant variance reduction.
Control variates for $H_{k:\ell}$ are proposed 
in \citet{craiu2022double}. They observe that $\mathbb{E}[h(X_t)-h(Y_t)]=0$ for all $t\geq 0$,
thus $\sum_{t\geq 0} \eta_t \{h(X_t)-h(Y_t)\}$ can be added to $H_{k:\ell}$, for any real sequence $(\eta_t)$, without modifying its expectation.
Optimization over the sequence $(\eta_t)$ can lead to a reduction in variance.
Other coupling-based
variance reduction strategies have been proposed for MCMC 
\citep{neal2001improving}, as well as techniques related to the Poisson equation 
\citep{andradottir1993variance,alexopoulos2022variance}.

\textbf{Finite moments.} The following result is taken from \citet{douc2022solving}.
It assumes that $\pi_0$ has bounded Radon--Nikodym derivative with respect to $\pi$. For example,
if $\pi$ is continuous and strictly positive on $\mathbb{R}^d$, then the assumption rules
out the choice of $\pi_0$ as a Dirac mass,
but it allows $\pi_0$ to be the Uniform distribution on any ball with positive radius.

\begin{theorem}
	Under Assumption \ref{asmpt:meetingmoments} with $\kappa>1$, i.e. $\mathbb{E}[\tau^\kappa]<\infty$,
	let $h\in L^{m}(\pi)$ for some $m>\kappa/(\kappa-1)$.
	Assume that $\pi_0$ is such that $d\pi_0/d\pi\leq M$ with $M<\infty$.
	Then for any $k,\ell,L$,
	the estimator $H_{k:\ell}$ in \eqref{eq:H_kellL} is unbiased: $\mathbb{E}[H_{k:\ell}] = \pi(h)$, and
	satisfies $\mathbb{E}[|H_{k:\ell}|^{p}]<\infty$ for $p\geq1$ such that $\frac{1}{p}>\frac{1}{m}+\frac{1}{\kappa}$.
	\label{thm:finitepmoments}
\end{theorem}

Finiteness of the first two moments
is particularly useful.
Finiteness of the variance of $H_{k:\ell}$ is sufficient to
validate the following classical construction of confidence intervals: generate
$C$ independent copies of $H_{k:\ell}$, compute their average $\hat{\mu}$ and
their standard deviation $\hat{\sigma}$, and an asymptotically (as $C\to\infty$) valid confidence
interval for $\pi(h)$ is given by $[\hat\mu+q_{\alpha/2}
\hat\sigma/\sqrt{C},\hat\mu+q_{1-\alpha/2} \hat\sigma/\sqrt{C}]$, where $q_s$
is the $s$-th quantile of the standard Normal distribution.  According to
Theorem \ref{thm:finitepmoments}, finiteness of second moments ($p=2$)
holds under 
a mild condition on $\pi_0$, the assumption that $\kappa>2$ and for all $h\in
L^m(\pi)$ such that $m>2\kappa/(\kappa-2)$. If $\tau$ has Geometric tails,
then $\kappa$ can be taken arbitrary large, and the result holds for all $h\in L^{2+\eta}(\pi)$ for any $\eta>0$.

\subsection{Unbiased signed measure\label{sec:unbiasedmcmc:signedmeasure}}

\begin{figure}
\begin{center}
  {\includegraphics[width=\textwidth]{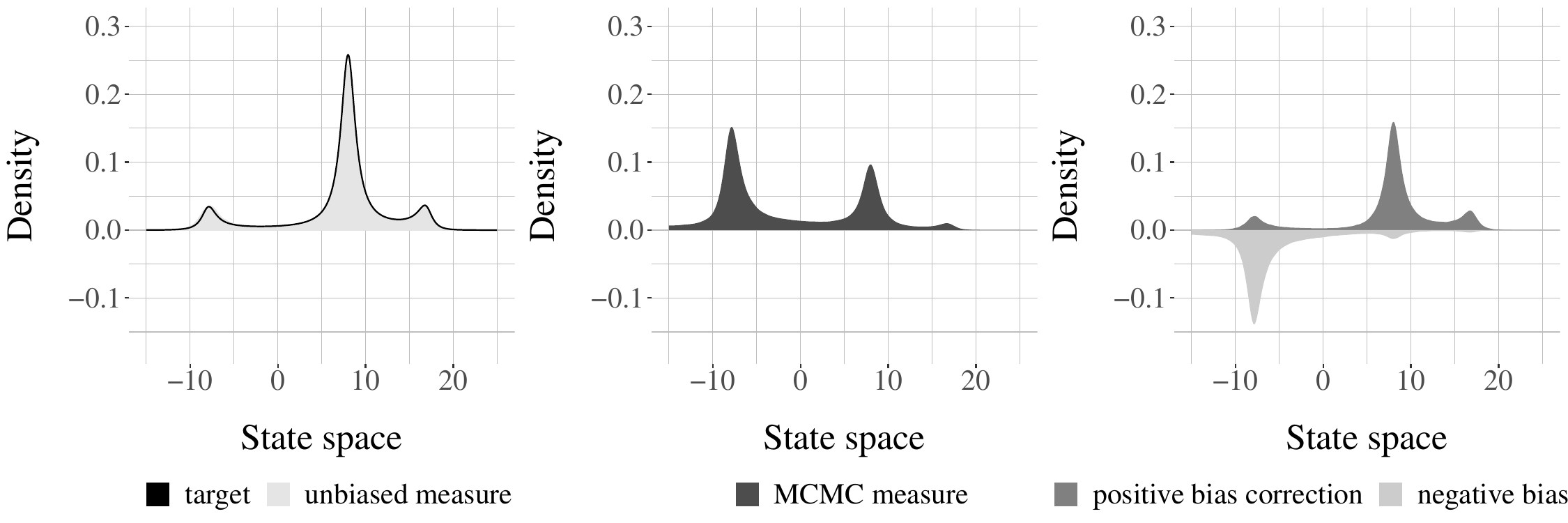}} 
  \caption{Unbiased MCMC (left) = MCMC (middle) + bias cancellation (right).
    The target density is the black curve on the left-most plot.
    On the right, the bias cancellation is made of a positive measure (darker grey) added to a negative measure (lighter grey).\label{fig:signedmeasures}}
\end{center}
\end{figure}

\textbf{Replacing function evaluations by Dirac masses.}
The empirical measure
\begin{equation}
  \hat{\pi}({\rm d}x)=\underbrace{\frac{1}{\ell-k+1}\sum_{t=k}^{\ell}\delta_{X_{t}}({\rm d}x)}_{\text{MCMC}}
  +\underbrace{\sum_{t=k+L}^{\tau-1}v_t(k,\ell,L)\left\{ \delta_{X_{t}}-\delta_{Y_{t-L}}\right\}({\rm d}x)}_{\text{bias cancellation}} ,\label{eq:pihat_kellL}
\end{equation}
is an unbiased approximation of $\pi$, where $v_t$ is defined in \eqref{eq:vtweight}. This is of the form $\sum_{n=1}^N \omega_n \delta_{Z_n}$ as in
\eqref{eq:signedmeas}, with $N=\max(0, \tau-(k+L))+(\ell-k+1)$, $Z_n$ are
states from either $(X_t)$ or $(Y_t)$ and $\omega_n$ are either
$(\ell-k+1)^{-1}$, or of the form $\pm v_n(k,\ell,L)$; in particular the
weights can be negative.  Figure \ref{fig:signedmeasures} represents the unbiased 
MCMC approximation, made of 
MCMC and bias cancellation components. This figure was created using kernel density estimation
from the weighted samples constituting the different elements in \eqref{eq:pihat_kellL}.

\textbf{Sub-sampling and negative weights.} We can sub-sample from the empirical measure in \eqref{eq:signedmeas}.  For
example, we can draw an index $I$ uniformly in $\{1,\ldots,N\}$ and return the
sample $Z_I$ with weight $N\omega_I$. Then for a class of functions $h$,
$N\omega_I h(Z_I)$ will have expectation equal to $\pi(h)$
\citep{douc2022solving}. We can also sample the index $I$ non-uniformly, with probabilities
$\xi_1,\ldots,\xi_N$ that depend on the atoms in \eqref{eq:signedmeas}, 
and the selected atom $Z_I$ is then weighted by $\xi_I^{-1}\omega_I$,
and we may repeat this selection multiple times to obtain a weighted sub-sample
from \eqref{eq:signedmeas} with a desired size.
Yet this does not produce a perfect sample due to the weights
being possibly negative. We can arbitrarily decrease the proportion of negative weights
in \eqref{eq:signedmeas} by
increasing the value of $k$, but we cannot make it zero. 
As a result, 
unbiased MCMC estimators $H_{k:\ell}$ can take values outside the range of the function $h$,
e.g. we
may obtain negative estimates of positive quantities.
There may not be any general solution to this problem: according to Lemma 2.1 in \citet{jacobthiery}
there is no algorithm that takes unbiased estimators of a nonnegative
quantity as input (and nothing else), and returns nonnegative unbiased estimators of that same quantity.

\subsection{Efficiency, cost and tuning\label{sec:unbiasedmcmc:efficiency}}

\textbf{Asymptotic equivalence with MCMC.} Theorem \ref{thm:finitepmoments} validates unbiased MCMC 
for the estimation of $\pi(h)$ but does not help for the comparison of its performance with standard MCMC,
or the choice of the tuning parameters $k,\ell,L$. 
The guiding principle in the tuning of $k,\ell,L$ is that a judicious choice will make unbiased MCMC
competitive with standard MCMC in terms of cost and variance.
Proposition 3 in \citet{joa2020} provides conditions under which the increase of either $k$ or $\ell-k$
results in variance reduction, and in particular the variance of $H_{k:\ell}$ is shown
to be asymptotically equivalent to the variance of standard MCMC estimators as $\ell\to\infty$.
The result is shown under weaker conditions in \citet{middleton2020unbiased}.
In the same spirit \citet{douc2022solving} provide the following CLT for $H_{k:\ell}$ as $\ell\to\infty$,
where the asymptotic variance is the same as for standard MCMC.
\begin{theorem}
	Under Assumption \ref{asmpt:meetingmoments} with $\kappa>1$, let $h\in L^{m}(\pi)$ for some $m>2\kappa/(\kappa-1)$. 
Then for any $k\in\mathbb{N}$,
\begin{equation}
  \sqrt{\ell-k+1}\left(H_{k:\ell}-\pi(h)\right) \overset{d}{\to}\text{Normal}(0,v(P,h)),
  \label{eq:cltunbiasedMCMC}
\end{equation}
as $\ell\to\infty$, where $v(P,h)$ is the 
asymptotic variance in the CLT for MCMC averages \eqref{eq:clt}.
\label{thm:cltunbiasedMCMC}
\end{theorem}
The asymptotic equivalence with regular MCMC as $\ell\to\infty$ should be
expected since the initialization bias vanishes
as $\ell\to\infty$. It can be seen in the form of the bias cancellation term in \eqref{eq:H_kellL}:
the sum is over $\max(0,\tau-L-k)$ terms (irrespective of $\ell$) and the
weights in \eqref{eq:vtweight} decrease as $(\ell-k)^{-1}$. Thus, the bias
cancellation term disappears when $\ell-k$ increases.
The cost of $H_{k:\ell}$  in \eqref{eq:costHkell}  
behaves as $\ell$ when $\ell\to\infty$. Hence, both cost and variance of $H_{k:\ell}$
are equivalent to those of MCMC as $\ell\to\infty$. 
By carefully choosing $k,\ell,L$ we can obtain unbiased MCMC estimators
with an efficiency close to that of MCMC. 

\begin{figure}
  \begin{center}
    {\includegraphics[width=.48\textwidth]{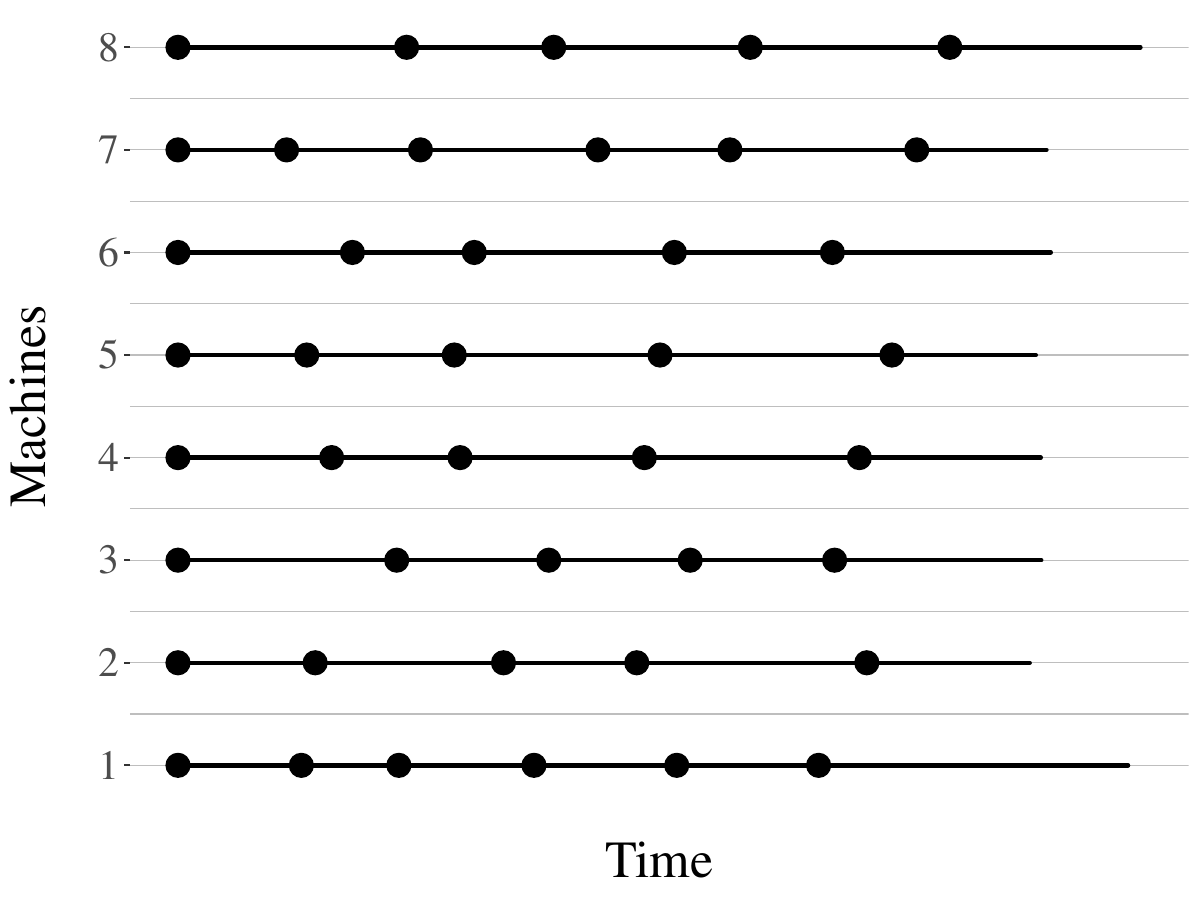}}\hfill
    {\includegraphics[width=.48\textwidth]{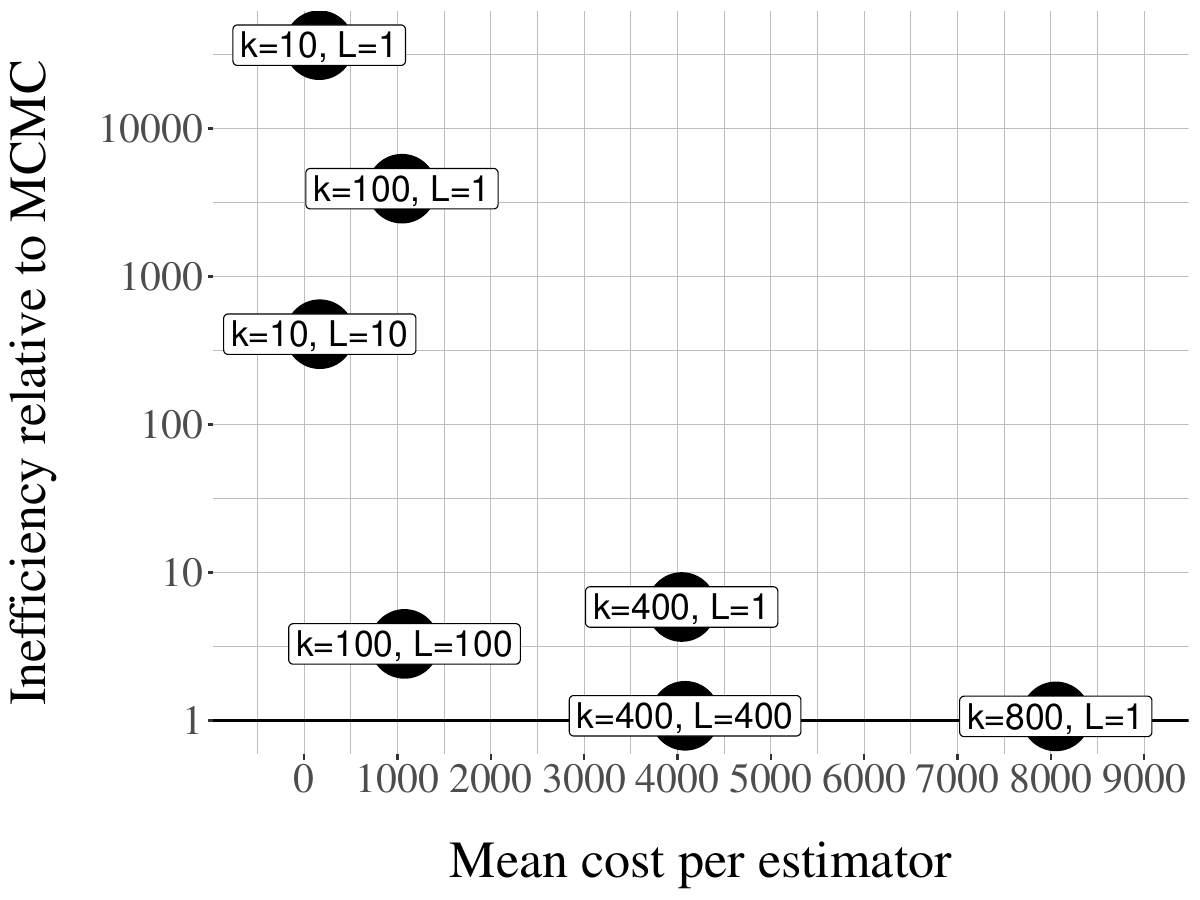}} 
    \caption{Left: chronology of the generation of
    estimators on parallel machines, with each dot representing the start of a run. Right: inefficiency of unbiased MCMC divided to the inefficiency of MCMC, versus mean cost of $H_{k:\ell}$,
      for different choices of $k$, and $L$ set to either $1$ or $k$, always with $\ell=10k$. The configuration $k=800,L=800$ is not shown as it would be overlaid with $k=800,L=1$.\label{fig:ctimeinef}
    }
  \end{center}
\end{figure}

\textbf{Cost and parallel computing.} Efficiency is not 
the only criterion when tuning unbiased MCMC.
Some users might prefer less efficient but cheaper estimators 
when enough parallel machines are available to produce them. 
Consider generating $C$ estimates on $M$ parallel machines. When $M \ll C$,
each machine produces many of the $C$ estimates. The speed-up of using $M$ parallel machines is then close
to linear in $M$. On the other hand, if $M\geq C$, each machine produces one estimate, and the user must wait for the longest run to complete.
Careful: running unbiased MCMC on $M\gg C$ machines and retaining the $C$ estimators that are first completed would introduce a bias, since the 
estimator is not independent of its cost.
Figure \ref{fig:ctimeinef} (left) illustrates the chronology
of the generation of independent estimators on parallel machines, with each machine producing a random number
of estimators within a given time period.
If each machine is tasked to produce a single estimator, the total time is exactly the
(random) cost of the longest run, which behaves in average as an increasing function of the number
of machines (see relevant discussions in \citet{wang2024unbiased}). For example if
$\tau$ has Geometric tails, the average maximum cost of unbiased MCMC behaves
as $\log(C)$.  Handling of budget constraints, such as hard or soft deadlines, on parallel machines 
is discussed in \citet{glynn1990bias,glynn1991analysis,joa2020}.

\textbf{Choice of length $\ell$.} We proceed to proposing
guidance for the tuning parameters.  First we simplify the choice
by recommending that $\ell$ is set as a large multiple of $k$, for example
$\ell=10k$. This is because a portion $k/\ell$ of iterations is simply
discarded in the construction of \eqref{eq:H_kellL}, and we would like to limit this
apparent waste.  Thus, we are left with the choice of $k$ and $L$; increasing $k$ will automatically 
increase $\ell$ and $\ell-k$, thus decreasing the magnitude of the weights $v_t$ in the bias cancellation term.

\textbf{Choice of burn-in $k$ and lag $L$.}
The bias cancellation is exactly zero in the event $\{\tau - L < k\}$.
By setting $k$ as a large quantile of $\tau-L$, we ensure that the event occurs with high probability.
The increase of the lag $L$, compared to the choice $L=1$ in \citet{glynn2014exact,joa2020}, is advocated in \citet{VanettiDoucet2020}.
From the expression of the weights in \eqref{eq:vtweight},
increasing $L$ decreases the weights in the bias cancellation term, and thus brings $H_{k:\ell}$ closer to regular MCMC.
Furthermore, setting $L=k$ leads to a minor increase of cost per estimator compared to $L=1$, and thus the efficiency is typically
improved, sometimes drastically.

\textbf{Concrete guideline.} In our experience, satisfactory tuning can be done as follows. 
First generate independent meeting times with lag $L=1$ (by lack of a better guess).
Then set $k$ as a large (e.g. $99\%$) quantile of $\tau - L$,
which is the number of coupled transitions up to the meeting time.
Finally, redefine $L=k$, and set $\ell=10k$.
Figure \ref{fig:ctimeinef} (right) shows how different choices of $k,L$ (always with $\ell=10k$)
lead to vastly different costs and inefficiencies, for the estimation of $\pi(h)$ with $h:x\mapsto x$. In the figure, the inefficiency of unbiased MCMC  is divided by the asymptotic variance $v(P,h)$
of MCMC, estimated using the method of Section \ref{sec:byproducts:asvar}. 
The relative inefficiency becomes close to one when $\ell=10k$ increases, and setting $L=k$  instead of $L=1$ is often worthwhile.

\section{Beyond the estimation of stationary expectations\label{sec:byproducts}}

Unbiased MCMC provides estimators of stationary expectations $\pi(h)$, 
and also helps in addressing questions of interest to MCMC users,
including non-asymptotic convergence diagnostics (Section \ref{sec:byproducts:ubounds}), 
and efficiency comparisons (Section \ref{sec:byproducts:asvar}).

\subsection{Convergence diagnostics\label{sec:byproducts:ubounds}}

\begin{figure}
  \begin{center}
    {\includegraphics[width=.48\textwidth]{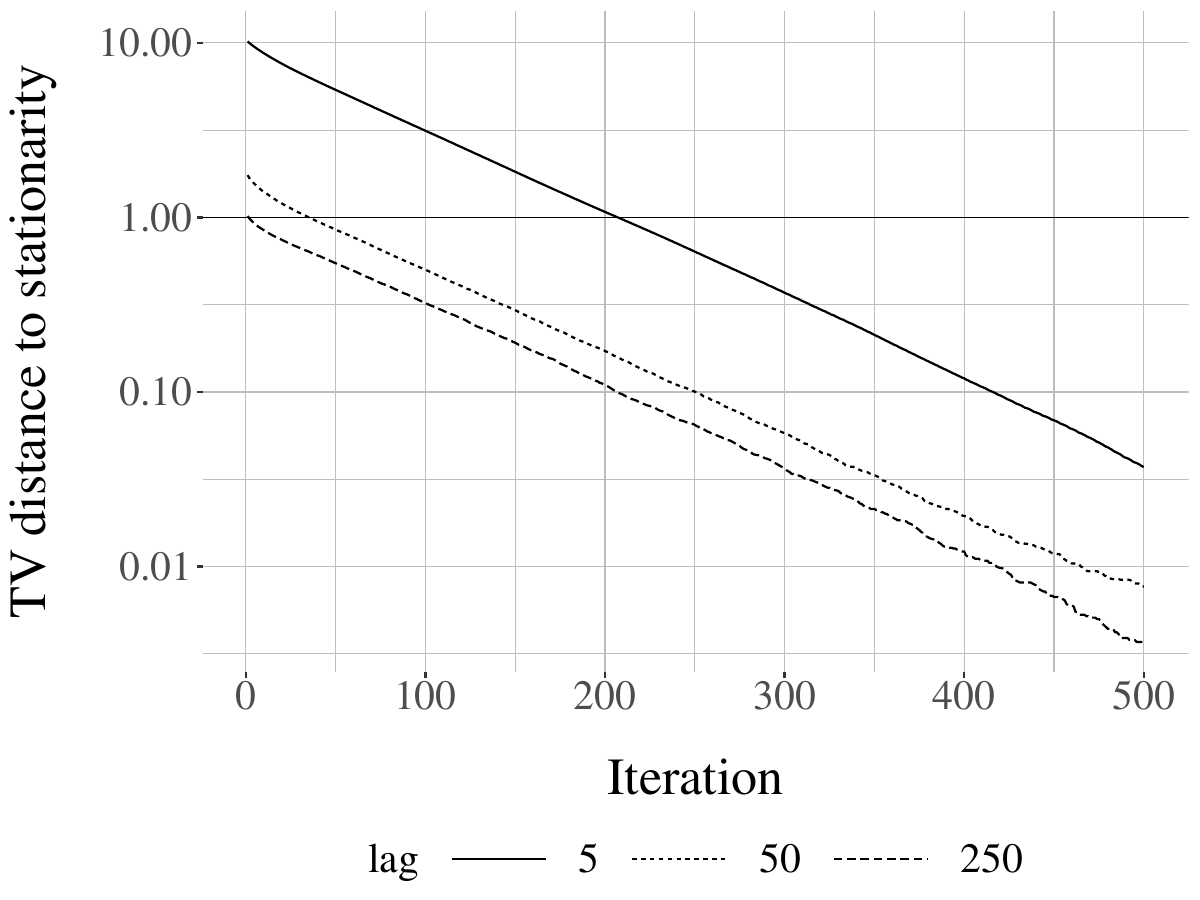}}\hfill
    {\includegraphics[width=.48\textwidth]{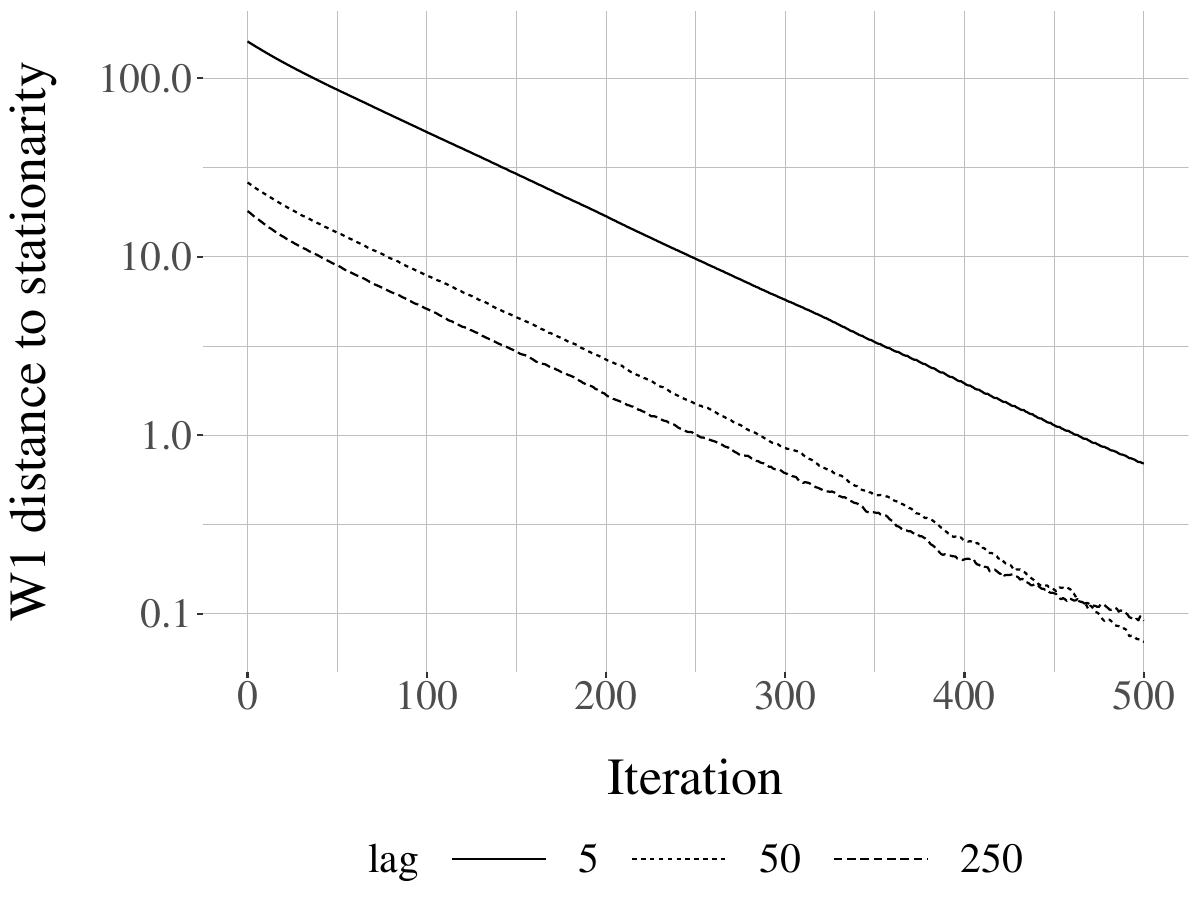}} 
    \caption{Left: upper bounds on $|\pi_{k}-\pi|_{\text{TV}}$ for different times $k$ on the x-axis, obtained
	    using \eqref{eq:upperboundtv} and $10^4$ independent copies of meeting times associated with lags $5$, $50$ and 250.  The total variation distance is always less than one. 
      Right: similar plot but for the 1-Wasserstein distance, obtained using \eqref{eq:upperboundw1}.
	    The y-axes are on logarithmic scale.}\label{fig:upperbounds}
  \end{center}
\end{figure}

\textbf{Total variation distance to stationarity.}  As a by-product of the unbiased estimator in \eqref{eq:HkL}
we can construct upper bounds on the total variation distance $|\pi_{k}-\pi|_{\text{TV}}$, for any finite $k$,
that can be estimated from samples of meeting times,
as first proposed in Section 6 of \citet{joa2020},
and improved with the use of $L>1$  in \citet{biswas2019estimating}, and with control variates in \citet{craiu2022double}.
A simple way of deriving such bounds is to write, for any $k\geq 0$,
\begin{equation}
|\pi_k - \pi|_{\text{TV}}= \inf_{(X_k,X)\in \Gamma(\pi_k,\pi)} \mathbb{E}[\mathds{1}(X_k\neq X)],
\end{equation}
where the infimum is over the set $\Gamma(\pi_k,\pi)$ of all pairs of variables $(X_k,X)$ where $X_k \sim \pi_k$ and $X \sim \pi$.
First use the triangle inequality to obtain an upper bound of the form $\sum_{j=1}^\infty |\pi_{k+jL} - \pi_{k+(j-1)L}|_{\text{TV}}$,
and then use the pair $(X_{k+jL},Y_{k+(j-1)L})$ generated by Algorithm \ref{alg:laggedchains} to obtain an upper bound $\mathbb{E}[\mathds{1}(X_{k+jL}\neq Y_{k+(j-1)L})]$ on each term in the sum.
Thus, a swap of expectation and limit yields
\begin{equation}
	 |\pi_{k}-\pi|_{\text{TV}}\leq\mathbb{E}[\sum_{j=1}^\infty \mathds{1}(X_{k+jL} \neq Y_{k+(j-1)L})] = \mathbb{E}[\max(0,\lceil(\tau-L-k)/L\rceil)].\label{eq:upperboundtv}
\end{equation}
The right-hand side is obtained by counting the indices $j\geq 1$ such that $k+jL$ is within $\{L,\ldots,\tau -1\}$. The  bounds can be estimated by replacing the expectation by an empirical average over $C$ independent 
meeting times $\tau_1,\ldots,\tau_C$, for any value of $k$. Thus one can run Algorithm \ref{alg:laggedchains}
with lag $L$ and $\ell=0$, $C$ times independently. The empirical upper bound
$C^{-1}\sum_{c=1}^C \max(0,\lceil(\tau_c-L-k)/L\rceil)$
is exactly zero
for all $k\geq \max_{c}\tau_c-L$, so it is enough to evaluate it at integers $k$ less than $\max_{c}\tau_c-L$.
Figure \ref{fig:upperbounds} (left) shows these bounds obtained for three different lags.
Increasing the lag tends to decrease the bounds, but the sharpness of the bounds also depends on the choice of coupling. To choose $L$, as before,
we can first generate meeting times with $L=1$, 
and then redefine $L$ as a large (e.g. $99\%$) empirical quantile of $\tau-L$, the number of coupled transitions leading to the meeting time.

\textbf{1-Wasserstein distance to stationarity.}
A similar reasoning leads to upper bounds on other distances that have a coupling representation.
\citet{biswas2019estimating} consider the 1-Wasserstein distance, defined as
\begin{equation}
  |\pi_k - \pi|_{W_1} = \inf_{(X_k,X)\in \Gamma(\pi_k,\pi)} \mathbb{E}[|X_k - X|],
\end{equation}
where $|x-y|$ represents a distance between $x$ and $y$ such as the Euclidean distance on $\mathbb{R}^d$.
With the same reasoning, first use the triangle inequality, and then employ any coupling of $\pi_{k+jL}$ and $\pi_{k+(j-1)L}$ for $j\geq 1$.
Thus, by running Algorithm \ref{alg:laggedchains} with lag $L$ and $\ell = 0$, assuming the validity of an exchange of expectation and limit, we obtain the upper bounds:
\begin{equation}
	\forall k\geq 0 \quad |\pi_{k}-\pi|_{W_1}\leq\mathbb{E}[\sum_{j=1}^{\lfloor(\tau - k  - 1)/L\rfloor} |X_{k+jL} - Y_{k+(j-1)L}|].\label{eq:upperboundw1}
\end{equation}
The sum on the right-hand side runs until $j$ such that $k+jL< \tau$, after which each term is zero.
Figure \ref{fig:upperbounds} (right) represents these bounds for three different lags.
As noted in \citet{papp2022bounds}, the $p$-th power of the $p$-Wasserstein distance between $\pi_k$ and $\pi$ can be upper bounded in the same way.

\textbf{Practical significance.} We stress the usefulness of \eqref{eq:upperboundtv} or \eqref{eq:upperboundw1} compared to the 
usual bounds encountered in the literature on Markov chains.
In continuous state spaces, the coupling inequality due to Wolfgang Doeblin 
\citep{lindvall2002lectures} reads: 
$|\pi_{k}-\pi|_{\text{TV}}\leq \mathbb{P}(\tau>k)$ for all $k\geq 0$, where the meeting time
is that of a pair of chains (without any lag), started from
$\pi_0$ and $\pi$, but MCMC users can rarely sample from $\pi$. 
In discrete state spaces,
one can also write $\max_x |P^k(x,\cdot)-\pi|_{\text{TV}} \leq
\max_{x,y}\mathbb{P}_{x,y}(\tau>k)$ where the meeting time corresponds to chains
started from states $x,y$ \citep[Corollary 5.3 in][]{levin2017markov}. Optimizing over the states $x,y$ 
could be computationally difficult.
In contrast, the upper bounds described above  
involve pairs of chains started from an arbitrary $\pi_0$.

\textbf{Limitation.} As a warning, the following describes a situation where the use of
\eqref{eq:upperboundtv} would fail to provide reliable bounds on
$|\pi_{k}-\pi|_{\text{TV}}$.  Suppose that the target $\pi$ is
multimodal, and that chains tend to get stuck in local modes. Assume 
further that the initial distribution $\pi_0$ puts its mass entirely in a local
mode of $\pi$. The user might then observe a small empirical average of the meeting
time, even after many independent runs. 
Yet the expectation of the meeting time could be much larger.
Indeed, there could be a small probability that one chain moves to a different
mode before meeting the second chain, and in that event, the meeting time could take 
large values, driving the expectation upward.  This is illustrated in
Section 5.1 of \citet{joa2020}. The risk is mitigated by specifying an
initial distribution $\pi_0$ that is spread out relative to the modes of $\pi$,
or by increasing the lag $L$ \citep{biswas2019estimating}.

\subsection{Efficiency comparison\label{sec:byproducts:asvar}}

Unbiased MCMC and its connection to the Poisson equation \citep{douc2022solving},
see Section \ref{sec:unbiasedmcmc:poisson},
lead to unbiased estimators of the asymptotic variance
$v(P,h)$ in the CLT \eqref{eq:clt}.

\textbf{Asymptotic variance through the Poisson equation.}
A standard way of establishing the CLT for Markov chain averages \citep[][Chapter 21]{douc2018MarkovChains} is to write
$\sum_{s=0}^{t-1}\{h(X_{s})-\pi(h)\} = \sum_{s=1}^{t}\left\{ g(X_{s})-Pg(X_{s-1})\right\} +g(X_{0})-g(X_{t})$,
where $g$ is a solution of the Poisson equation \eqref{eq:poisson},
and then to observe that $\left\{ g(X_{s})-Pg(X_{s-1})\right\}_{s\geq 1}$
forms a martingale difference sequence.
The CLT for martingale
difference sequences and some routine calculations yield
\begin{equation}
	v(P,h)=\mathbb{E}_\pi[\{g(X_{1})-Pg(X_{0})\}^{2}] =  2 \underbrace{\pi(\{h-\pi(h)\}g)}_{\text{(a)}} - \underbrace{(\pi(h^{2}) - \pi(h)^{2})}_{\text{(b)}}.
\label{eq:vPhfishyrepresentation}
\end{equation}

From the above representation of $v(P,h)$,
\citet{douc2022solving} combine
unbiased estimators $G$ of evaluations of $g$, from \eqref{eq:Gxy} in Section \ref{sec:unbiasedmcmc:poisson},
with unbiased MCMC approximations $\hat{\pi}$ of $\pi$ as in \eqref{eq:pihat_kellL}, to deliver estimators $\hat{v}(P,h)$
with $\mathbb{E}[\hat{v}(P,h)] = v(P,h)$. The estimator in its simplest form goes as follows. First, run two independent unbiased MCMC estimators $\hat{\pi}^{(j)}$, for $j=1,2$, as in \eqref{eq:pihat_kellL}. Using $\hat{\pi}^{(1)},\hat{\pi}^{(2)}$ we can estimate without bias the term $(b)$ in \eqref{eq:vPhfishyrepresentation} by 
\begin{equation}
  (B) = (\hat{\pi}^{(1)}(h^2)+\hat{\pi}^{(2)}(h^2))/2 - \hat{\pi}^{(1)}(h)\hat{\pi}^{(2)}(h).
\end{equation}

To estimate $(a)$ in \eqref{eq:vPhfishyrepresentation}, recall from Section \ref{sec:unbiasedmcmc:poisson} that, for any $x$, we can run coupled chains to obtain an unbiased estimator $G(x,y)$ in \eqref{eq:Gxy} of $g(x,y)$ in \eqref{eq:gxy}, for any fixed choice of $y\in\mathbb{X}$. Then, write $\hat{\pi}^{(1)}$ in the form $\sum_{n=1}^N \omega_n \delta_{Z_n}$. Draw $I$ uniformly in $\{1,\ldots,N\}$, and run coupled chains started from $Z_I$ and $y$ to obtain $G(Z_I,y)$. Conditioning on $\hat\pi^{(1)},\hat\pi^{(2)}$ and $I$, we have
\begin{equation}
\mathbb{E}\left[N\omega_I G(Z_I,y)\left(h(Z_I) - \hat{\pi}^{(2)}(h)\right)\vert \;\hat\pi^{(1)},\hat\pi^{(2)},I\right] = N\omega_I g(Z_I,y)\left(h(Z_I) - \hat{\pi}^{(2)}(h)\right).
\end{equation}
By further averaging out $I$ we obtain $\hat\pi^{(1)}(g(\cdot,y)h) - \hat\pi^{(2)}(h)\hat\pi^{(1)}(g(\cdot,y))$,
and since $\hat{\pi}^{(j)}$ for $j=1,2$ are unbiased, we see that 
\begin{equation}
  (A) = N\omega_IG(Z_I,y)\left(h(Z_I) - \hat{\pi}^{(2)}(h)\right),
\end{equation}
is an unbiased estimator of $(a)$ in \eqref{eq:vPhfishyrepresentation}. Thus, $\hat{v}(P,h) = 2 (A) - (B)$ is an unbiased estimator of $v(P,h)$.
This estimator can be improved in various ways, for instance by drawing multiple copies of $I$ and estimating the solution of the Poisson equation at the corresponding states $Z_I$. We refer the reader to \citet{douc2022solving} for more details on such estimators.


\textbf{Practical significance.}
The quantity $v(P,h)$ measures the efficiency of the underlying MCMC algorithm,
and thus constitutes
a reference value for the efficiency of unbiased MCMC, as seen in Section \ref{sec:unbiasedmcmc:efficiency}. Access to the unbiased estimators of $v(P,h)$  enables efficiency comparisons between standard and unbiased MCMC, such as that represented in
Figure \ref{fig:ctimeinef} (right), without ever performing long MCMC runs.
Comparisons can also be done between MCMC algorithms, 
since the estimators are unbiased for the asymptotic variance, and are not upper bounds as in Section \ref{sec:byproducts:ubounds}. 

Under assumptions guaranteeing the existence of a finite variance
for the unbiased estimator $\hat{v}(P,h)$ of $v(P,h)$,
averages of $C$ independent copies would converge at 
the Monte Carlo rate. This compares favorably to classical estimators
of $v(P,h)$ based on long runs.
Indeed, commonly-used estimators of $v(P,h)$, 
such as batch means and spectral variance estimators,
converge at a sub-Monte Carlo rate, e.g. $T^{-2/3}$  for batch means \citep{flegal2010batch}.
A limitation of the unbiased estimation strategy is that it requires 
a successful coupling of the algorithm under consideration, whereas
classical estimators only require trajectories of the chain.



\section{Design of successful coupling of MCMC algorithms \label{sec:designcoupling}}

To implement unbiased MCMC, users
need to design a successful
coupling of their MCMC algorithm.  
Focusing on Markovian couplings, 
this amounts to constructing a coupled transition $\bar{P}$
to plug in Algorithm \ref{alg:laggedchains} and for which 
Assumption \ref{asmpt:meetingmoments} is satisfied.
Concretely, we need to be able to sample $(X,Y) \sim \bar{P}((x,y),\cdot)$,
where $(x,y)$ represent the current positions of the chains,
such that 1) $X\sim P(x,\cdot)$ and $Y\sim P(y,\cdot)$, and 
2) the resulting chains meet as quickly as possible.
In Section \ref{sec:designcoupling:ingredients} we review 
the more basic task of coupling random variables,
before dealing with MCMC transitions in Section \ref{sec:designcoupling:combine}.
References to realistic examples
are provided in Section \ref{sec:designcoupling:examples}.  

\subsection{Couplings of random variables\label{sec:designcoupling:ingredients}}

\textbf{Maximal couplings.} 
A coupling of $(X,Y)$ with $X\sim p$ and $Y\sim q$ is \emph{maximal} if 
$\mathbb{P}(X=Y)$ is maximal and thus equal 
to $1-|p-q|_{\text{TV}}$.
There may be more than one maximal coupling.  Algorithm
\ref{alg:submaximalcoupling} is a modification by \citet{GerberLee2020}
of the $\gamma$-coupling of \citet{johnson1998coupling} with an extra 
parameter $\eta\in(0,1]$. The scheme requires 
samples from $p$ and $q$, and
evaluations of the ratio of their densities. The probability
of $\{X=Y\}$
is maximal only when $\eta=1$. However, the cost of
running Algorithm \ref{alg:submaximalcoupling}, which contains a while loop,
has a variance that goes to infinity when 
$\eta=1$ and when $|p-q|_\text{TV}$ goes to zero.  
With $\eta<1$, the coupling is sub-maximal, but the variance of the cost is upper bounded uniformly over $p$ and $q$. 
Under  Algorithm \ref{alg:submaximalcoupling}, conditionally on $Y$ being generated in step 2.(b), $X$ is independent of $Y$. 

\begin{algorithm}
\begin{enumerate}
\item Sample $X\sim p$.
\item Sample $W\sim\text{Uniform}(0,1)$. 
\begin{enumerate}
\item If $W\leq \min(\eta,q(X)/p(X))$, set $Y=X$.
\item Otherwise sample $Y^{\star}\sim q$ and $W^{\star}\sim\text{Uniform}(0,1)$
until $W^{\star}>\eta p(Y^{\star})/q(Y^{\star})$,
\item[] and set $Y=Y^{\star}$.
\end{enumerate}
\item Return $(X,Y)$.
\end{enumerate}
\caption{Sampling a coupling of $p$ and $q$, with parameter $\eta\in (0,1]$.
  The coupling maximizes $\mathbb{P}(X=Y)$ when $\eta=1$, but the variance
  of the cost is bounded when $\eta<1$.\label{alg:submaximalcoupling}}
\end{algorithm}

\begin{figure}
  \begin{center}
    {\includegraphics[width=.48\textwidth]{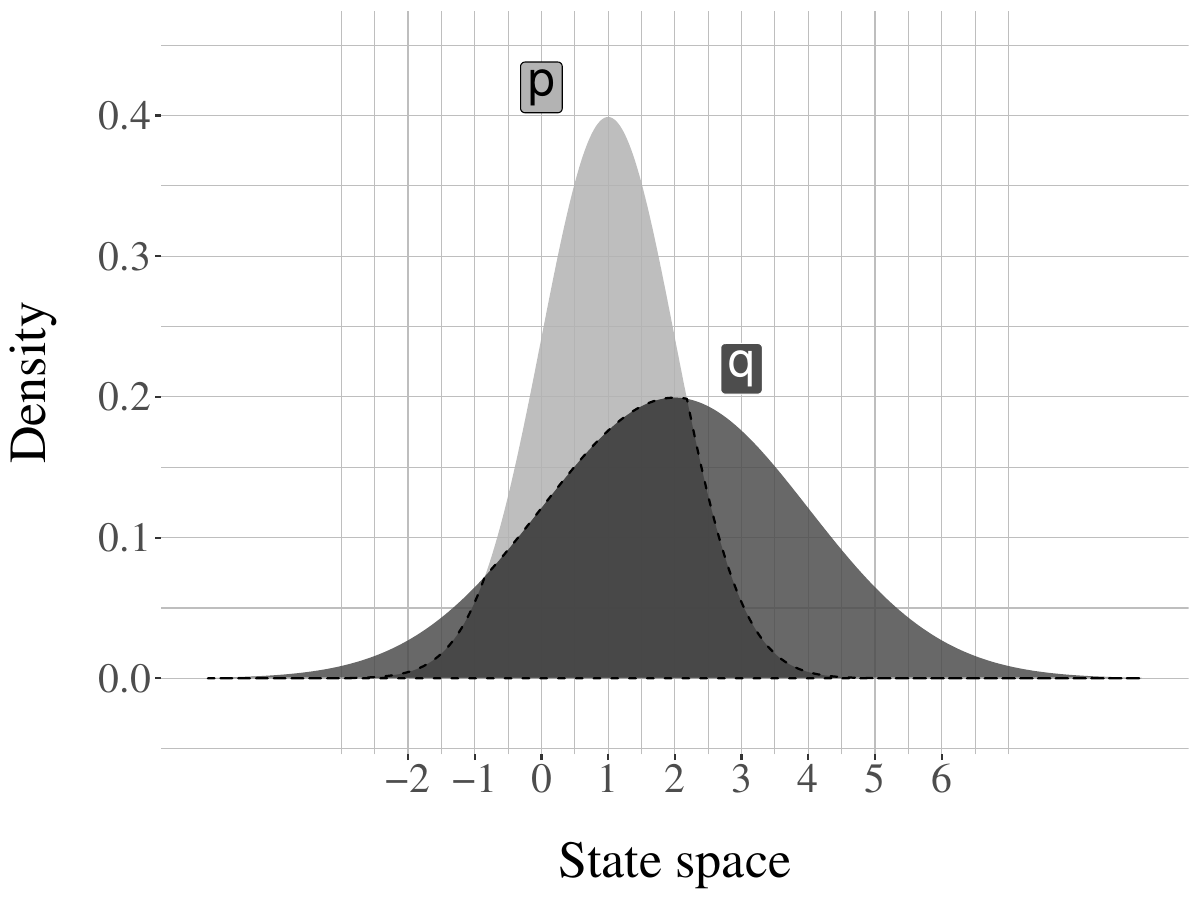}}
    {\includegraphics[width=.48\textwidth]{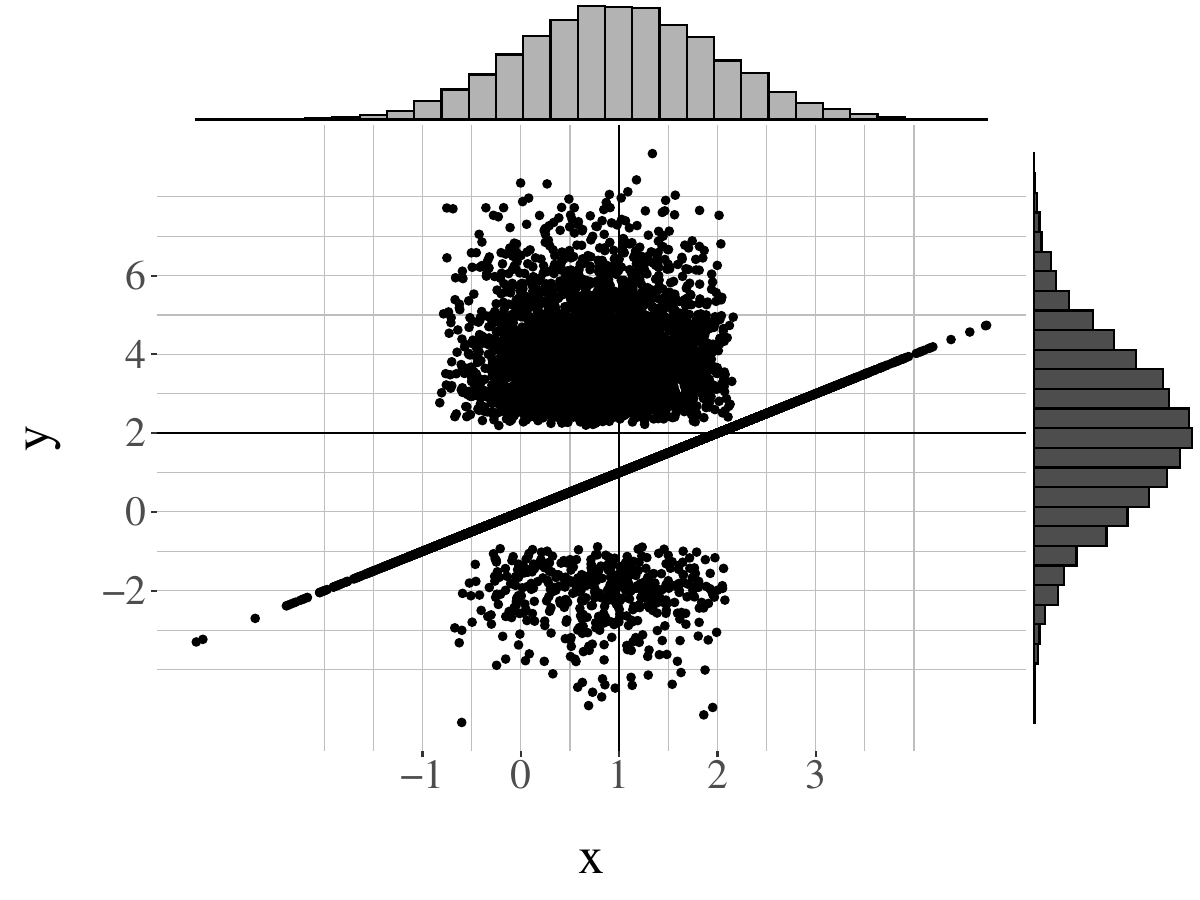}}
    \caption{Left: distributions $p$ and $q$ defined as $\text{Normal}(\mu_1,\sigma_1^2)$, with $\mu_1=1,\sigma_1=1$,
    and $\text{Normal}(\mu_2,\sigma_2^2)$, with $\mu_2=2,\sigma_2=2$. Right: samples $(x_i,y_i)$ following a maximal coupling of $p$ and $q$ using Algorithm \ref{alg:submaximalcoupling}, shown as a scatter plot with marginal histograms.\label{fig:couplings}}
  \end{center}
\end{figure}

Algorithm \ref{alg:maximalcoupling-mixture} samples the same pairs $(X,Y)$ 
as Algorithm \ref{alg:submaximalcoupling} with $\eta=1$,
but via a mixture representation; see e.g. \citet{devroye1990coupled,bremaud1992maximal}, and Chapter 1 in \citet{thorisson2000coupling}.
Algorithm \ref{alg:maximalcoupling-mixture} is applicable when $\int \min(p(x),q(x))dx$
can be computed, and when $\tilde{p}$ and $\tilde{q}$ defined on line 3 can be sampled from.
Note that $|p-q|_{\text{TV}}=1-\int \min(p(x),q(x))dx$.
Its appeal is that its cost is deterministic. Maximal couplings are illustrated in Figure \ref{fig:couplings} in the case 
of two univariate Normal distributions. The figure shows the two probability density functions (left), and a sample of pairs $(x_i,y_i)$ generated using Algorithm \ref{alg:submaximalcoupling} with $\eta=1$. Notably, some pairs $(x_i,y_i)$ are such that $x_i=y_i$.

\begin{algorithm}
\begin{enumerate}
\item Draw $U\sim\text{Uniform}(0,1)$ and compute $c = \int \min(p(x),q(x))dx$.
\item If $U\leq c$, 
draw $X\sim \nu$ with $\nu:x\mapsto \min(p(x),q(x))/c$, and
set $Y=X$.
\item Otherwise, draw $X$ from $\tilde{p}:x\mapsto (p(x)-\min(p(x),q(x)))/(1-c)$,
\item[] and independently  draw $Y$ from $\tilde{q}:x\mapsto (q(x)-\min(p(x),q(x)))/(1-c)$.
\item Return $(X,Y)$.
\end{enumerate}
\caption{Sampling a maximal coupling of $p$ and $q$ via a mixture.\label{alg:maximalcoupling-mixture}}
\end{algorithm}

\textbf{Synchronous couplings.}
The above algorithms generate $X$ and $Y$ independently, conditionally on the event that the draw of $X$ is not accepted for $Y$.
Thus they revert to the independent coupling  when $|p-q|_{\text{TV}}$ is close to one.
A natural way of introducing dependencies among random variables
is to use \emph{common random numbers} to generate them. 
In one dimension, if $X$ has quantile function $F_p^-$ and $Y$ has quantile function $F_q^-$,
then one can sample $U\sim\text{Uniform}(0,1)$ and set $X=F_p^-(U)$
and $Y=F_q^-(U)$. The resulting joint distribution
minimizes $\mathbb{E}[(X-Y)^2]$, or 
equivalently, maximizes $\mathbb{C}\text{ov}(X,Y)$ \citep{glasserman1992some}. 
Hence, using common random numbers can correspond to an \emph{optimal transport} coupling \citep{villani2008optimal}.
We will see below that these couplings can help in devising contracting chains.

\textbf{Reflection couplings.}
Spherically symmetric random variables
are invariant by reflections with respect to planes
passing through their center. 
From this observation, reflection couplings can be designed
for e.g.  Normal or
Student distributions with means $\mu_1$,$\mu_2$ and equal variance
$\Sigma$, in any dimension.  The sample $Y$ is defined by reflection of $X$ with respect to the hyperplane 
bisecting the segment $(\mu_1,\mu_2)$. The resulting coupling is synchronous 
in all directions orthogonal to the difference $(\mu_1 - \mu_2)$, 
along which it is anti-synchronous: $X-\mu_1$ and $Y-\mu_2$ either point toward each other,
or face opposite directions.
\citet{bou2020coupling} propose a coupling that
both maximizes $\mathbb{P}(X=Y)$
and reverts to a reflection coupling
in the event $\{X\neq Y\}$, described in 
Algorithm \ref{alg:reflmaxcoupling} and with a deterministic cost.
Figure \ref{fig:Rcode} includes an implementation.

\begin{algorithm}
  \begin{enumerate}
    \item Let $\Delta=\Sigma^{-1/2}(\mu_{1}-\mu_{2})$ and $e=\Delta/|\Delta|$ where $|\cdot|$ is the $L_2$ norm.
    \item Sample $\dot{X}\sim s$, and $W\sim\text{Uniform}(0,1)$.
    \item If $s(\dot{X})W\leq s(\dot{X}+\Delta)$, set $\dot{Y}=\dot{X}+\Delta$.
    \item Else set $\dot{Y}=\dot{X}-2(e^{T}\dot{X})e$.
    \item Set $X=\Sigma^{1/2}\dot{X}+\mu_{1},Y=\Sigma^{1/2}\dot{Y}+\mu_{2}$, and return $(X,Y)$.
  \end{enumerate}
\caption{A coupling for distributions 
$p$ and $q$ obtained from a common spherically symmetric distribution with density $s(\cdot)$, rescaled
with a common covariance $\Sigma$, and shifted by $\mu_1$ and $\mu_2$ respectively.
For example $p=\text{Normal}(\mu_{1},\Sigma)$
and $q=\text{Normal}(\mu_{2},\Sigma)$ if $s=\text{Normal}(0,I)$.\label{alg:reflmaxcoupling}}
\end{algorithm}


\subsection{Coupling MCMC transitions\label{sec:designcoupling:combine}}

\textbf{Coupling the constituents of a transition.}
MCMC algorithms describe how to obtain $X_t$ conditional on $X_{t-1}=x$ 
through a succession of steps. 
With MRTH (e.g. Figure \ref{fig:Rcode}), a proposal $X^\star$ is sampled 
from a transition $q(x,\cdot)$ (step 1),
then $U$ is sampled from Uniform$(0,1)$
and $X_t$ is set to $X^\star$ if $U<\pi(X^\star)q(X^\star,x)/\pi(x)q(x,X^\star)$,
or to $x$ otherwise (step 2).
Couplings of the entire transition can be 
constructed by coupling each step, e.g. 
coupling proposals $X^\star \sim q(x,\cdot)$ and  $Y^\star \sim q(y,\cdot)$,
and then coupling the Uniforms employed for accepting or rejecting the proposals.
For example \citet{johnson1998coupling} uses maximal couplings as in Algorithm \ref{alg:submaximalcoupling} for the proposals, and a common Uniform for acceptance. \citet{wang2021maximal} refine the
coupling of the Uniforms to maximize the probability of $\{X_t = Y_t\}$. 
\citet{o2021metropolis} show that all couplings of MRTH transitions
can be obtained by certain stepwise couplings.

\textbf{Contracting before meeting.}
For some MCMC algorithms
it may be possible to sample from a
maximal coupling of $P(x,\cdot)$ and $P(y,\cdot)$
\citep[e.g.][for MRTH]{wang2021maximal}. However,
even the maximal probability of $\{X=Y\}$, which is $1-|P(x,\cdot)-P(y,\cdot)|_{\text{TV}}$,
is very small unless $x$ and $y$ are close to one another.
Thus, the aim is first to bring the chains closer,  
so that they may then have a decent chance to meet.
A coupling of $P$ may alternate between 
different strategies depending on the current states $x$ and $y$:
for example one can employ a \emph{contractive coupling} of $P$ (as described below) if $|x-y|$ is
large and a maximal coupling of $P$ if $|x-y|$ is small.
\citet{eberle2016} refers to such alternation as \emph{mixed couplings}.

\textbf{Contractive couplings.}
In the context of Markov chains,
the transition $X_t \sim P(X_{t-1},\cdot)$ can be represented 
as $X_{t} = \psi(X_{t-1},U_t)$,
where $U_t$ is a source of randomness and $\psi$ is a deterministic function.
Then  the \emph{synchronous coupling}  refers
to the computation of $X_t=\psi(X_{t-1},U_t)$ and $Y_t=\psi(Y_{t-1},U_t)$ using 
the same random variable $U_t$ at time $t$.
It has long been observed
that synchronous couplings of MCMC algorithms can be contractive
\citep{johnson1996studying,neal1999circularly,neal2001improving},
in the sense that the generated chains tend to get closer to one another.
Assuming strong convexity of the potential function,
it is known that common noise terms
result in contraction for Langevin diffusions 
\citep[pages 22-23 in][]{villani2008optimal}, for unadjusted Langevin \citep[e.g. Appendix A in][]{wibisono2018sampling},
and for Hamiltonian Monte Carlo \citep[e.g.][]{mangoubi2017rapid}
for at least some tuning parameters.
Contraction from synchronous couplings has been observed 
in the context of Gibbs samplers, e.g. 
in \citet{biswas2021couplingbased} for linear regression with horseshoe priors,
and \citet{atchadewang2023} for regression with spike-and-slab priors,
as well as for the preconditioned Crank--Nicholson algorithm in \citet{agapiou2018unbiased}.

Reflection couplings were introduced to analyze Brownian motions
on Euclidean spaces, leading to the smallest possible 
meeting times \citep{lindvall1986coupling,hsu2013maximal}.
Reflections were later employed to obtain contraction for various processes:
\citet{eberle2016} for a class of diffusion processes,
\citet{eberle2019langevin} for Langevin dynamics,
\cite{bou2020coupling} for Hamiltonian Monte Carlo, for example.
\citet{joa2020} observe good performance
of Algorithm \ref{alg:reflmaxcoupling} for random walk proposals in MRTH,
on spherical Normal distributions as the dimension increases.
\citet{papp2022new} establish this formally and propose another coupling, termed
\emph{gradient common random number coupling}, which is shown to
work optimally for a class of target distributions.

\textbf{Mixing different MCMC transitions to enable meetings.}
For an MCMC algorithm with transition $P_1$, it may be possible to design a contractive coupling $\bar{P}_1$
without being able to induce meetings.
For example with Hamiltonian Monte Carlo (HMC), 
contraction can result from
the use of common momentum variables \citep[e.g][]{mangoubi2017rapid}.
However, to obtain meetings would require 
pairs of momentum variables such that two
Hamiltonian trajectories, propagated
with these momentum variables, would end up at the same final position
\citep[see Figure 1 in][]{bou2023mixing}. A way to 
bypass this difficulty is to introduce another transition $P_2$ 
with a coupling $\bar{P}_2$ that induces meetings when chains are close.
\citet{heng2019unbiased} then propose
to use a chain with transition defined by the mixture $w_1 P_1 + w_2 P_2$, with $w_1+w_2=1$:
with probability $w_1$, the chain evolves with $P_1$ (e.g. HMC), and otherwise with $P_2$ (e.g. random walk MRTH).
A coupling of such mixture of transitions can be defined as a mixture of the coupled transitions, 
$w_1 \bar{P}_1 + w_2 \bar{P}_2$.
The resulting chains may contract thanks to $\bar{P}_1$, e.g. HMC with common random numbers,
and have a chance to meet thanks to $\bar{P}_2$, e.g. MRTH with maximally coupled proposals. Careful:
it would not be legal to  
employ $\bar{P}_1$ when the chains are distant and $\bar{P}_2$ when they are close,
as this would violate
the marginal constraint that each chain evolves according to  $w_1 P_1 + w_2 P_2$.

\subsection{References to couplings of realistic MCMC algorithms\label{sec:designcoupling:examples}}

Successful couplings have been developed for a number of 
popular MCMC algorithms.

\textbf{Discrete state spaces.}
Convergence diagnostics are 
challenging on discrete spaces, for which 
visualization is difficult; there, unbiased MCMC could be particularly useful. 
\citet{joa2020} present a coupling of 
the Gibbs sampler studied in \citet{yang2016}
for Bayesian variable selection in high dimension.
\citet{nguyen2022many} couple Gibbs samplers to perform
Bayesian data clustering, where the states are partitions
of finite sets.
\citet{kelly2021lagged} couple
MCMC samplers for phylogenetic inference,
where the state space is that of discrete tree topologies
along with parameters and latent variables.

\textbf{Particle filtering and importance sampling.}
Conditional particle filters for smoothing in state space
models are coupled in 
\citet{jacob2018smoothing}.
\citet{lee2020coupled} extend the methodology
and propose a detailed study of the meeting times.
Particle marginal Metropolis--Hastings for Bayesian
inference in state space models \citep{andrieu2010particle} is
coupled in \citet{middleton2020unbiased}.
Particle independent Metropolis--Hastings
is coupled in \citep{middleton2019unbiased},
with the curious implication that
the bias of self-normalized importance sampling estimators can be removed in finite time;
and likewise for general sequential Monte Carlo samplers. 
\citet{ruiz2021unbiased} couple variants of iterated sampling importance resampling
to fit variational auto-encoders.

\textbf{Gradient-based MCMC.}
\citet{heng2019unbiased,xu2021couplings} consider couplings of
simple variants of Hamiltonian Monte
Carlo with applications to logistic regression and log-Gaussian Cox point processes
in non-trivial dimensions. Reflection couplings as in Figure \ref{fig:Rcode} or Algorithm \ref{alg:reflmaxcoupling} can be directly used for Langevin
Monte Carlo.
\citet{corenflos2023debiasing} propose couplings
of some piecewise deterministic MCMC algorithms
such as the bouncy particle sampler \citep{bouchard2018bouncy}. 

\textbf{Tempering.}
\citet{joa2020} couple a parallel tempering version of
a Gibbs sampler for the Ising model.
\citet{zhu2020minimax} consider
coupled simulated tempering for sparse canonical correlation analysis.


\section{Comments and outstanding questions\label{sec:discussion}}

\subsection{Possible uses beyond parallel computing}

Access to unbiased signed measures approximating
the target $\pi$ facilitates parallel computing: 
instead of long chains, unbiased MCMC users rely on
large numbers of independent runs. The lack of bias has other appeals.

\textbf{Expectation inside optimization loops.}
Iterative optimization methods may require the approximation of an
expectation at each iteration, which should preferably be unbiased
to prevent accumulation of bias over the iterations \citep[e.g.][]{tadic2011asymptotic}. 
The usefulness of unbiased
MCMC is investigated for a Monte Carlo Expectation-Maximization scheme in
\citet{chen2018blind}, and for stochastic gradient optimization for variational
auto-encoders in \citet{ruiz2021unbiased}.

\textbf{Leveraging the statistical toolbox.}
Access to independent unbiased estimators of an expectation $\pi(h)$ enables
the direct use of the statistical toolbox.
For example one can readily replace empirical averages
by more robust estimators of expectations.
\citet{nguyen2022many} consider trimmed means.
One could naturally employ median-of-means estimators
\citep{lugosimendelson2019}, empirical risk optimizers \citep{sun2024need}
or estimators based on self-normalized sums \citep{MinskerNdaoud} to aggregate unbiased MCMC estimators $\hat{\pi}(h)$ that have two finite moments
under conditions stated in Theorem \ref{thm:finitepmoments}.
Unbiased estimators can also be plugged into the framework 
of multi-arm bandits, for example to identify 
the algorithm with minimal asymptotic variance
among a collection of MCMC algorithms. 
One could view each algorithm as
an arm, and each unbiased estimator $\hat{v}(P,h)$ of an asymptotic variance $v(P,h)$ 
as an observed loss.
Then, \emph{best arm identification} techniques \citep{audibert2010best}
can be used to find, as efficiently as possible, the arm associated with the smallest expected loss.

Consider expectations with respect to a
distribution on $\mathbb{X}_1\times \mathbb{X}_2$ defined as
$\pi_{12}(x_1,x_2)=\pi_1(x_1)\pi_2(x_2|x_1)$.  Suppose that $\pi_1(x_1)$ can be
evaluated up to a normalizing constant $Z_1$, and that $\pi_2(x_2|x_1)$ can be
evaluated up to a normalizing constant $Z_2(x_1)$, which is not constant with
respect to $x_1$. The unnormalized density of $\pi_{12}(x_1,x_2)$ involves the
term $Z_2(x_1)$. If $Z_2(x_1)$ cannot be evaluated, then standard MCMC
algorithms cannot be implemented \citep{plummer2015cuts}. The setting occurs commonly
in various parts of data analysis
\citep[e.g.][]{blocker2013potential,liu2009modularization,jacob2017better,rainforth2018nesting}.
Such nested distributions can be conveniently approximated with an unbiased MCMC estimator at each level \citep[see examples in][]{joa2020,rischard2018unbiased}, as a simple consequence of the law of total expectation.
Relatedly \citet{wang2022unbiased} consider the 
problem of estimating a nonlinear function $g$ of an expectation $\pi(h)$. They develop
generic unbiased estimators of $g(\pi(h))$ by combining unbiased MCMC with unbiased
multilevel Monte Carlo \citep{blanchet2019unbiased}.

\subsection{Applicability}

There exists a world, of a size to be determined, between standard MCMC and perfect sampling,
where unbiased estimators can be obtained but not exact samples \citep{glynn2016exact}.
Successful 
couplings of MCMC algorithms open a door to that world. Currently, such construction is endeavored algorithm-by-algorithm,
by mixing ingredients such as maximal couplings,
common random numbers and reflections.
There is no guarantee that such \emph{ad hoc} constructions can always
be found. For some ergodic chains,
there exist couplings such that $|\pi_t - \pi_0|_\text{TV} = \mathbb{P}(\tau > t)$ \citep[e.g.][]{pitman1976coupling}, 
but these may not often be implementable in settings of relevance for MCMC practitioners. 
It is however possible to plug arbitrary Markov transitions into mixtures of kernels,
as described in Section \ref{sec:designcoupling:combine},
or in an SMC sampler, and then to remove its bias via a generic coupling
of particle independent Metropolis--Hastings \citep{middleton2019unbiased}.

A successful coupling of inhomogeneous Markov transitions, e.g. for adaptive MCMC algorithms, remains elusive. For a stochastic process $(X_t)$ with marginals converging to $\pi$ in the sense that $\pi(h) = \lim_{t\to\infty} \mathbb{E}[h(X_t)]$, assuming that $(Y_t)$ is a copy of $(X_t)$ such that
$\sum_{t\geq 1}\mathbb{E}\left[|h(X_t) - h(Y_{t-1})|\right]$
is finite,
then $\pi(h)$ has the representation 
\begin{equation}\label{eq:repres:pi}
\pi(h) = \mathbb{E}[h(X_0) + \sum_{t\geq 1}\left(h(X_t) - h(Y_{t-1})\right)].\end{equation}
Many adaptive MCMC algorithms are known to have converging marginals \citep{andrieu-thoms-2008,atchade-etal-2011}. In principle the debiasing device could be applied to \eqref{eq:repres:pi}, but it is unclear how to construct a faithful coupling of $(X_t)$ and $(Y_t)$ 
that would lead to an unbiased estimator with a finite computing time. Random truncation techniques as in Section \ref{sec:unbiasedmcmc:telescop} could be used, but good performance would depend on a choice of truncation variable
that may require detailed knowledge of $(X_t)$.


Links to code repositories and complementary information can be found
on the companion website at \url{https://pierrejacob.quarto.pub/unbiased-mcmc}.

\paragraph*{Acknowledgments.} The authors are grateful to the editors of the second edition of the Handbook of Markov chain Monte Carlo, and to 
Pieter Jan Motmans, El Mahdi Khribch and anonymous reviewers who provided useful feedback on earlier versions of this manuscript.


\bibliographystyle{apalike} 
\bibliography{handbook}

\end{document}